\documentclass[letter,usenatbib]{mn2e}
\usepackage{epsfig}
\usepackage{afterpage}
\def\gtrsim{~\rlap{$>$}{\lower 1.0ex\hbox{$\sim$}}}

\title[Variations in 24~$\mu$m morphologies in SINGS]
    {Variations in 24~$\mu$m morphologies among galaxies in the
    {\it Spitzer} Infrared Nearby Galaxies Survey: New insights
    into the Hubble sequence}
\author[G. J. Bendo et al.]
    {G. J. Bendo,$^1$, D. Calzetti,$^2$, C. W. Engelbracht,$^3$
    R. C. Kennicutt, Jr.,$^4$, M. J. 
    \newauthor Meyer,$^5$ M. D. Thornley,$^6$
    F. Walter,$^7$ D. A. Dale,$^8$ A. Li,$^9$ and E. J. Murphy$^{10}$ \\
    $^1$Astrophysics Group, Imperial College, Blackett Laboratory,
        Prince Consort Road, London SW7 2AZ, United Kingdom\\
    $^2$Department of Astronomy, University of Massachusetts, 
        LGRT-B 254, 710 North Pleasant Street, Amherst, MA 01002, USA\\
    $^3$Steward Observatory, University of Arizona, 933 North 
        Cherry Avenue, Tucson, AZ 85721, USA\\
    $^4$Institute of Astronomy, University of Cambridge, 
        Cambridge CB3 0HA, United Kingdom\\
    $^5$Space Telescope Science Institute, 3700 San Martin Drive, 
        Baltimore, MD 21218, USA\\
    $^6$Department of Physics \& Astronomy, Bucknell University, 
        Lewisburg, PA 17837, USA\\
    $^7$Max-Planck-Institut f\"ur Astronomie, K\"onigstuhl 17, 
        D-69117 Heidelberg, Germany\\
    $^8$Department of Physics and Astronomy, University of Wyoming, 
        Laramie, WY 82071, USA\\
    $^9$Department of Physics and Astronomy, University of 
        Missouri, Columbia, MO 65211, USA\\
    $^{10}$Department of Astronomy, Yale University, P.O. Box 208101, 
        New Haven, CT 06520, USA}
\date{}
\pagerange{\pageref{firstpage}--\pageref{lastpage}}
\pubyear{}

\begin{document}
\label{firstpage}
\maketitle

\begin{abstract}
To study the distribution of star formation and dust emission within
nearby galaxies, we measured five morphological parameters in the 3.6
and 24~$\mu$m wave bands for 65 galaxies in the {\it Spitzer} Infrared
Nearby Galaxies Survey (SINGS) and 8 galaxies that were
serendipitously observed by SINGS.  The morphological parameters
demonstrate strong variations along the Hubble sequence, including
statistically significant differences between S0/a-Sab and Sc-Sd
galaxies.  Early-type galaxies are generally found to be compact,
centralized, symmetric sources in the 24~$\mu$m band, while late-type
galaxies are generally found to be extended, asymmetric sources.
These results suggest that the processes that increase the real or
apparent sizes of galaxies' bulges also lead to more centralized
24~$\mu$m dust emission.  Several phenomena, such as strong nuclear
star formation, Seyfert activity, or outer ring structures, may cause
galaxies to deviate from the general morphological trends observed at
24~$\mu$m.  We also note that the 24~$\mu$m morphologies of Sdm-Im
galaxies are quite varied, with some objects appearing very compact
and symmetric while others appear diffuse and asymmetric.  These
variations reflect the wide variation in star formation in irregular
galaxies as observed at other wavelengths.  The variations in the
24~$\mu$m morphological parameters across the Hubble sequence mirror
many of the morphological trends seen in other tracers of the ISM and
in stellar emission.  However, the 24~$\mu$m morphological parameters
for the galaxies in this sample do not match the morphological
parameters measured in the stellar wave bands.  This implies that the
distribution of dust emission is related to but not equivalent to the
distribution of stellar emission.
\end{abstract}


\section{Introduction}

The variations in the integrated properties of the interstellar medium
(ISM) and integrated star formation activity along the Hubble sequence
is clearly defined.  \citet{rh94} and references therein have shown
that the total gas surface density and the ratio of gas mass to total
mass increases when proceeding along the Hubble sequence from
elliptical galaxies through early-type spiral galaxies (galaxies with
large bulges and tightly-wound arms) to late-type spiral galaxies
(galaxies with small or negligible bulges and loosely-wound arms).
\citet{k98a} and references therein have used various tracers of
star formation normalized by total stellar content to establish the
general increase in star formation activity when proceeding from
elliptical galaxies to late-type spiral galaxies along the Hubble
sequence.  Moreover, all of these trends show significant differences
in the interstellar content and star formation activity between early-
and late-type spiral galaxies.

However, few investigations have carefully studied whether the spatial
distribution of the ISM or star formation varies among spiral
galaxies, particularly between Sa and Sd galaxies.  \citet{hk83} show
that the radial distributions of H{\small II} regions are broader in
late-type than in early-type spiral galaxies, but their sample only
contains 37 galaxies and includes no galaxies earlier than Sb.  Later
H$\alpha$ surveys appear to contradict these results; a larger
H$\alpha$ survey by \citep{dghhc01} found no variations in the radial
extent of the H$\alpha$ emission normalized by the spatial extent of
the I-band emission, and another H$\alpha$ survey by \citet{khc06}
found no variations in the ratio of the H$\alpha$/R-band scale
lengths.  \citet{betal02} and \citet{tacgde04} did demonstrate that
the dust emission was more extended in late-type spiral galaxies than
in early-type spiral galaxies, but the data in \citet{betal02} did not
include all of the galaxy emission, and the sample in \citet{tacgde04}
contained very few galaxies with Hubble types earlier than Sb.
\citet{pafw04} qualitatively show a similar variation in the
non-stellar 8.0~$\mu$m morphology of galaxies across the Hubble
sequence but did not apply a quantitative analysis.  \citet{yetal95}
did demonstrate strong variations in the CO emission across the Hubble
sequence, but the observations consisted mostly of a series of single
low-resolution pointed observations along the major axes of the target
galaxies.  In contrast to these other observations, \citet{tacgde04}
observed that the spatial extent of H{\small I} emission did not vary
with Hubble type, at least between Sb and Sd galaxies.

Variations in the distribution of the ISM and star formation among
spiral galaxies could have fundamental implications for the stellar
evolution and the evolution of structure within galaxies.  Moreover,
such variations may provide clues to the processes that form galaxies'
disks and bulges.  For example, if the ISM is more centrally
concentrated in early-type spiral galaxies than in late-type spiral
galaxies, this may point to merger phenomena as having formed the
bulges of early-type spiral galaxies and causing gas to fall into the
galaxies' nuclei \citep[e.g.][]{s98}, or it may indicate that the
centres of early-type galaxies contain pseudobulges, which form
through secular processes that funnel gas into the nuclei and trigger
nuclear star formation activity \citep[e.g.][]{kj-kr04}.  Also note
that the distribution of dust emission may be related to other galaxy
properties.  For example, \citet{detal07} demonstrated that the
clumpiness and central concentration of 24~$\mu$m emission was linked
to the dust temperatures and the ratio of infrared to ultraviolet
luminosity.  Links between the distribution of dust and galaxy
morphology may therefore have implications for how other galaxy
properties vary along the Hubble sequence as well.

The {\it Spitzer} Infrared Nearby Galaxies Survey
\citep[SINGS;][]{ketal03} recently completed mid- and far-infrared
observations of nearby galaxies that represent a cross section of
galaxies with different Hubble types, luminosities, and
infrared/optical ratios.  Additionally, this survey detected many
serendipitous sources, including nearby galaxies not included in the
SINGS sample.  In this paper, we use the 24~$\mu$m {\it Spitzer} data
to understand how the distribution of this component of dust emission
varies among nearby galaxies.  For comparison with stellar
morphologies, we also examine the morphologies in the 3.6~$\mu$m
images, which trace stellar emission that mostly originates from
evolved stellar populations \citep[e.g.][]{letal03}.

In general, infrared dust emission can be used as an approximate
tracer of both star formation and interstellar dust mass.
Technically, dust emission represents the fraction of the total
stellar radiation that is absorbed by dust.  However, since hot young
stars frequently produce a significant fraction of the total
bolometric luminosity within star formation regions and since dust is
concentrated near star formation regions, dust emission is an
effective star formation tracer.  This has been demonstrated recently
with comparisons between {\it Spitzer Space Telescope} 8 - 160~$\mu$m
data and H$\alpha$ or ultraviolet emission in nearby galaxies
including M~81 \citep{getal04,petal06} and M~51 \citep{ckbetal05}.
The 24~$\mu$m wave band, which originated from both transiently-heated
very small grain and grains that are in thermal equilibrium at
$\sim100$~K, appears to be the best star formation tracer of all the
IRAC and MIPS bands.  While the 24~$\mu$m band is particularly
sensitive to dust heating \citep{dhcsk01,ld01} and may therefore be
less than ideal for tracing dust mass, the high spatial resolution
(6~arcsec) and signal-to-noise ratios in the 24~$\mu$m data
are superior to any other mid- or far-infrared data available from
{\it Spitzer} or elsewhere.  The 24~$\mu$m data are therefore the best
data available at this time for studying the spatial distribution of
dust in nearby galaxies.  Note that the 24~$\mu$m emission may not
trace star formation very well in some dwarf galaxies with stochastic
star formation \citep[e.g.][]{cwbetal05, cetal06} and active galactic
nuclei may be responsible for significant amounts of dust heating in
the nuclei of some galaxies \citep[e.g.][for an extreme
example]{betal06a}, but the 24~$\mu$m emission can still provide some
information on star formation even in these cases.

In Section~\ref{s_data}, we discuss the sample used for the analysis,
the {\it Spitzer} observations with the Infrared Array Camera
\citep[IRAC;][]{fetal04} and the Multiband Imaging Photometer for {\it
Spitzer} \citep[MIPS;][]{ryeetal04}, the data reduction and the
parameterization of the morphologies of the target galaxies.  We rely
on the morphological parameters defined by \citet{c03} and
\citet{lpm04} as well as the half-light radius of the 24~$\mu$m
emission normalized by optical radius to study the morphologies.  In
Section~\ref{s_test}, we examine the effects of distance and
inclination angle on these parameters, and we derive corrections for
dealing with these effects where necessary.  In
Section~\ref{s_morphparam_ttrend}, we study how the morphological
parameters vary along the Hubble sequence, with particular emphasis
placed on searching for variation in morphology from S0/a-Sab through
to Sc-Scd spiral galaxies, and we also select galaxies with
representative or peculiar 24~$\mu$m morphologies for their Hubble
types.  We briefly compare the 24~$\mu$m morphologies of barred and
unbarred spiral galaxies in Section~\ref{s_morphparam_bar}.  We then
discuss in Section~\ref{s_discuss} how the results compare to previous
studies of the ISM and optical morphologies and what the implications
are for interpreting the Hubble sequence.  A summary of the results is
presented in Section~\ref{s_conclusions}.

\section{Data} \label{s_data}

\subsection{Sample} \label{s_data_sample}

\begin{table}
\begin{center}
\renewcommand{\thefootnote}{\alph{footnote}}
\caption{Basic Properties of the Sample Galaxies from SINGS
    \label{t_sample_sings}}
\begin{tabular}{@{}lccc@{}}
\hline
    Name &
    Hubble &
    Size of Optical &
    Distance\\
    &
    Type\footnotemark[1] &
    Disc (arcmin)\footnotemark[1] &
    (Mpc)\footnotemark[2] \\
\hline
DDO 53 &     Im &                
    $1.5 \times 1.3$\footnotemark[3] &      3.5 \\
Ho I &        IAB(s)m &           
    $3.6 \times 3.6$\footnotemark[4] &      3.5 \\
Ho II &       Im &                $7.9 \times 6.3$ &     3.5 \\
IC 2574 &     SAB(s)m &           $13.2 \times 5.4$ &    3.5 \\
IC 4710 &     SB(s)m &            $3.6 \times 2.8$ &     8.5 \\
Mrk 33 &      Im pec &            
    $1.0 \times 0.9$\footnotemark[4] &      21.7 \\
NGC 24 &      SA(s)c &            $5.8 \times 1.3$ &     8.2 \\
NGC 337 &     SB(s)d &            $2.9 \times 1.8$ &     24.7 \\
NGC 628 &     SA(s)c &            $10.5 \times 9.5$ &    11.4 \\
NGC 855 &     E &                 $2.6 \times 1.0$ &     9.6 \\
NGC 925 &     SAB(s)d &           $10.5 \times 5.9$ &    10.1 \\
NGC 1097 \footnotemark[5] &
    SB(s)b &            $9.3 \times 6.3$ &     16.9 \\
NGC 1266 &    (R')SB0(rs) pec &   
    $1.5 \times 1.0$\footnotemark[4] &      31.3 \\
NGC 1291 &    (R)SB(s)0/a &       
    $9.8 \times 8.1$\footnotemark[4] &      9.7 \\
NGC 1316 &    SAB(s)0 pec &       $12.0 \times 8.5$ &    26.3 \\
NGC 1377 &    S0 &                $1.8 \times 0.9$ &     24.4 \\
NGC 1482 &    SA0$^+$ pec &       $2.5 \times 1.4$ &     22.0 \\
NGC 1512 &    SB(r)a &		  $8.9 \times 5.6$ &     10.4 \\
NGC 1566 &    SAB(s)bc &	  $8.3 \times 6.6$ &     18.0 \\
NGC 1705 &    SA0$^-$ pec &	  $1.9 \times 1.4$ &     5.8 \\
NGC 2403 &    SAB(s)cd &	  $21.9 \times 12.3$ &   3.5 \\
NGC 2798 &    SB(s)a pec &	  $2.6 \times 1.0$ &     24.7 \\
NGC 2841 &    SA(r)b &		  $8.1 \times 3.5$ &     9.8 \\
NGC 2915 &    I0 &		  $1.9 \times 1.0$ &     2.7 \\
NGC 2976 &    SAc pec &		  $5.9 \times 2.7$ &     3.5 \\
NGC 3031 &    SA(s)ab &		  $26.9 \times 14.1$ &   3.5 \\
NGC 3049 &    SB(rs)ab &	  $2.2 \times 1.4$ &     19.6 \\
NGC 3184 &    SAB(rs)cd &	  $7.4 \times 6.9$ &     8.6 \\
NGC 3190 &    SA(s)a pec &        $4.4 \times 1.5$ &     17.4 \\
NGC 3198 &    SB(rs)c &		  $8.5 \times 3.3$ &     9.8 \\
NGC 3265 &    E &		  $1.3 \times 1.0$ &     20.0 \\
NGC 3351 &    SB(r)b &		  $7.4 \times 5.0$ &     9.3 \\
NGC 3521 &    SAB(rs)bc &         $11.0 \times 5.1$ &    9.0 \\
NGC 3621 &    SA(s)d &		  $12.3 \times 7.1$ &    6.2 \\
NGC 3627 &    SAB(s)b &		  $9.1 \times 4.2$ &     8.9 \\
NGC 3773 &    SA0 &		  $1.2 \times 1.0$ &     12.9 \\
NGC 3938 &    SA(s)c &		  
    $5.4 \times 4.9$\footnotemark[4] &      12.2 \\
NGC 4125 &    E6 pec &		  $5.8 \times 3.2$ &     21.4 \\
NGC 4236 &    SB(s)dm &		  $21.9 \times 7.2$ &    3.5 \\
NGC 4254 &    SA(s)c &		  
    $5.4 \times 4.7$\footnotemark[4] &      20.0 \\
NGC 4321 &    SAB(s)bc &	  $7.4 \times 6.3$ &     20.0 \\
NGC 4450 &    SA(s)ab &		  $5.2 \times 3.9$ &     20.0 \\
NGC 4536 &    SAB(rs)bc &	  $7.6 \times 3.2$ &     25.0 \\
NGC 4559 &    SAB(rs)cd &	  $10.7 \times 4.4$ &    11.6 \\
NGC 4569 &    SAB(rs)ab &	  $9.5 \times 4.4$ &     20.0 \\
NGC 4579 &    SAB(rs)b &	  $5.9 \times 4.7$ &     20.0 \\
NGC 4594 &    SA(s)a &		  $8.7 \times 3.5$ &     13.7 \\
NGC 4625 &    SAB(rs)m pec &      
    $2.2 \times 1.9$\footnotemark[4] &      9.5 \\
NGC 4631 &    SB(s)d &		  $15.5 \times 2.7$ &    9.0 \\
NGC 4725 &    SAB(r)ab pec &	  $10.7 \times 7.6$ &    17.1 \\
NGC 4736 &    (R)SA(r)ab &        $11.2 \times 9.1$ &    5.3 \\
NGC 4826 &    (R)SA(rs)ab &	  $10.0 \times 5.4$ &    5.6 \\
NGC 5033 &    SA(s)c &		  $10.7 \times 5.0$ &    13.3 \\
NGC 5055 &    SA(rs)bc &	  $12.6 \times 7.2$ &    8.2 \\
NGC 5194\footnotemark[6] &	  
    SA(s)bc pec &    $11.2 \times 6.9$ &    8.2 \\
NGC 5398 &    (R')SB(s)dm pec &	  $2.8 \times 1.7$ &     15.0 \\
NGC 5408 &    IB(s)m &		  
    $1.6 \times 0.8$\footnotemark[4] &      4.5 \\
NGC 5474 &    SA(s)cd pec &	  
    $4.8 \times 4.3$\footnotemark[4] &      6.9 \\
NGC 5713 &    SAB(rs)bc pec &	  $2.8 \times 2.5$ &     26.6 \\
NGC 5866 &    SA0$^+$ &		  $4.7 \times 1.9$ &     12.5 \\
\hline
\end{tabular}
\end{center}
\end{table}

\begin{table}
\begin{center}
\renewcommand{\thefootnote}{\alph{footnote}}
\contcaption{}
\begin{tabular}{@{}lccc@{}}
\hline
    Name &
    Hubble &
    Size of Optical &
    Distance\\
    &
    Type\footnotemark[1] &
    Disc (arcmin)\footnotemark[1] &
    (Mpc)\footnotemark[2] \\
\hline
NGC 6822 &    IB(s)m &            
    $15.5 \times 13.5$\footnotemark[3] &    0.6 \\ 
NGC 6946 &    SAB(rs)cd &	  
    $11.5 \times 9.8$\footnotemark[4] &     5.5 \\
NGC 7331 &    SA(s)b &		  $10.5 \times 3.7$ &    15.7 \\
NGC 7552 &    (R')SB(s)ab &       $3.4 \times 2.7$ &     22.3 \\
NGC 7793 &    SA(s)d &            $9.3 \times 6.3$ &     3.2 \\
\hline
\end{tabular}
\end{center}
$^a$ These data are taken from RC3.  The optical disc is the size of the 
     D$_{25}$ isophote.\\
$^b$ These data are taken from \citet{ketal03}.  The distances are
     calculated using H$_0$=70 km s$^{-1}$ Mpc$^{-1}$ and the
     systematic velocities in the Nearby Galaxies Catalog \citep{t88}.
     Objects in the Virgo Cluster were set to a distance of 20~Mpc, and
     objects in the M81 Group were set to a distance of 3.5~Mpc.\\
$^c$ Significant dust emission is detected outside the optical
     discs of these galaxies.  Morphological parameters were therefore 
     measured in the larger regions described in Table~\ref{t_regspecial}.  See
     Section~\ref{s_data_paramdef} for more details.\\
$^d$ RC3 does not provide position angles for the major axes of
     these galaxies, so morphological parameters are measured within 
     circular regions with radii equal to the length of the major axes. See
     Section~\ref{s_data_paramdef} for more details.\\
$^e$ The optical disk of NGC~1097A is excluded from the analysis.
$^f$ A 2~arcmin region centered on NGC~5195 is excluded from the analysis.
\end{table}

The sample is selected from the SINGS sample of galaxies as well as
additional sources that were serendipitously observed in the SINGS
MIPS observations.  For a description of the SINGS sample, see
\citet{ketal03}.

A few SINGS sources were not suitable for this analysis.  We excluded
sources with surface brightnesses that did not exceed three times the
background noise in any MIPS wave bands.  This step not
only excludes non-detections but also galaxies where the primary
source of 24~$\mu$m emission may be starlight.  The dwarf irregular
galaxies DDO~154, DDO~165, Ho~IX, M81 Dwarf A, and M81 Dwarf B as well
as the elliptical galaxies NGC~584 and NGC~4552 are thus excluded from
the sample.  NGC~1404 was excluded because the only significant source
of emission within the optical disc at 70 and 160~$\mu$m is a point
source to the northeast of the nucleus that may not be physically
associated with the galaxy.  NGC~5195 (M51b) was excluded because a
significant fraction of the emission within its optical disc
originates from NGC ~5194 (M51a).  NGC~3034 (M82) was excluded because the
source saturated the 24~$\mu$m image.  The resulting sample contains
65 galaxies.

\begin{table}
\begin{center}
\renewcommand{\thefootnote}{\alph{footnote}}
\caption{Basic Properties of the Serendipitous Sample Galaxies}
\label{t_sample_seren}
\begin{tabular}{@{}lccc@{}}
\hline
    Name  &
    Hubble &
    Size of Optical &
    Distance \\
    &
    Type\footnotemark[1] &
    Disc (arcmin)\footnotemark[1] &
    (Mpc)\footnotemark[2]\\
\hline
IC 3583 &     IBm &               $2.2 \times 1.1$ &    20.0 \\
NGC 586 &     SA(s)a &            $1.6 \times 0.8$ &    28.9 \\
NGC 1317 &    SAB(r)a &           $2.8 \times 2.4$ &    25.7 \\
NGC 1510 &    SA0 pec &           $1.3 \times 0.7$ &    11.2 \\
NGC 3185 &    (R)SB(r)a &         $2.3 \times 1.6$ &    16.4 \\
NGC 3187 &    SB(s)c pec &        
    $3.0 \times 1.3$\footnotemark[3] &    21.4 \\
NGC 4533 &    SAd &               $2.1 \times 0.4$ &    23.9 \\
NGC 4618 &    SB(rs)m &           $4.2 \times 3.4$ &    8.6 \\
\hline
\end{tabular}
\end{center}
$^a$ These data are taken from RC3.  The optical disc is given as the
    major and minor axes of the D$_25$ isophote.\\
$^b$ To maintain consistency with \citet{ketal03}, the distances are
     calculated using H$_0$=70 km s$^{-1}$ Mpc$^{-1}$ and the
     systematic velocities in either the Nearby Galaxies Catalog
     \citep{t88} or, if those velocities were not available, the
     equivalent velocities from RC3.  IC 3583 was set at the distance
     used by Kennicutt et al. for the Virgo Cluster (20~Mpc).\\
$^c$ RC3 does not provide a position angle for the major axis of NGC
     3187, so morphological parameters are measured within a circular
     region with a radius equal to the length of the major axes. See
     Section~\ref{s_data_paramdef} for more details.
\end{table}
\renewcommand{\thefootnote}{\arabic{footnote}}

Although a large number of serendipitous sources were detected in this
survey, we placed several constraints on the serendipitous sources
that could be used for this study.  First, we only used sources listed
in the Third Reference Catalogue of Bright Galaxies
\citep[RC3;][]{ddcbpf91} that are within 2500 km/s (the approximate
maximum redshift for the SINGS sample).  Next, we required the sources
to have surface brightnesses that exceeded three times the background
noise in all MIPS wave bands.  Finally, we excluded NGC~2799 because
the optical disc of the galaxy contains significant emission from the
point spread function of the emission from NGC~2798.  The final sample
of serendipitous sources contains 8 galaxies.

Basic properties of the sample galaxies are listed in
Tables~\ref{t_sample_sings} and \ref{t_sample_seren}.  Note that this
sample is not an unbiased sample of nearby galaxies.  The SINGS sample
is not chosen to be a representative cross-section of nearby galaxies
but instead is chosen to contain nearby galaxies with a broad range of
properties, including a broad range of infrared luminosities and a
broad range of optical to infrared luminosity ratios.  Some of the
galaxies that were included in SINGS to completely sample this colour
and luminosity space may be unusual compared to the majority of nearby
galaxies.  The serendipitous sources tend to be late-type galaxies
that are physically associated with the SINGS galaxies, so they also
may not necessarily be representative of nearby galaxies in general.
As a consequence, any trends versus Hubble type for this sample may be
noisier than for an unbiased sample of galaxies such as a
distance-limited sample.  Nonetheless, since the sample used in this
paper still contains many average nearby galaxies, the sample should
still be useful for probing the infrared morphologies of nearby
galaxies.

\subsection{Observations and Data Reduction}

The 3.6 and 24~$\mu$m data were taken with {\it Spitzer} as part of
the SINGS legacy project.  The 3.6~$\mu$m IRAC observations for each
object consist of either a series of 5~arcmin $\times$ 5~arcmin
individual frames taken in either a mosaic or a single field dither
pattern.  The 24~$\mu$m MIPS observations comprise two scan maps for
each target.  Each object is observed twice in each wave band so as to
identify and remove transient phenomena, particularly asteroids.  The
FWHM of the PSFs, as stated in the Spitzer Observer's Manual
\citep{sscmanual06}\footnote[11]{http://ssc.spitzer.caltech.edu/documents/som/},
are 1.6~arcsec at 3.6~$\mu$m and 6~arcsec at
24~$\mu$m.  Details on the observations can be found in the
documentation for the SINGS fourth data delivery
\citep{sings06}\footnote[12]{Available at
http://ssc.spitzer.caltech.edu/legacy/singshistory.html}.

The IRAC data are processed using a SINGS IRAC pipeline which combines
multiple frames of data using a drizzle technique.  A description of
the technique is presented in \citet{retal06}.  The MIPS data were
processed using the MIPS Data Analysis Tools version 3.06
\citep{getal05} along with additional tools to remove zodiacal light
emission in the 24~$\mu$m data.  Additional details are presented in
\citet{betal06b}.  Note that the plate scales of the 24~$\mu$m data
used in this paper are 1.5 arcsec.  Full details are available in the
SINGS documentation for the fourth data delivery \citep{sings06}.

In the NGC~6822, some diffuse emission from the galaxy may have been
subtracted by the software that subtracts the zodiacal light emission
from the 24~$\mu$m data.  Although this effect has not been
quantified, we expect the 24~$\mu$m emission in NGC~6822 to be
dominated by bright, compact sources.  We therefore still include
NGC~6822 in this analysis, but the results from the 24~$\mu$m data should
be interpreted cautiously.

\subsection{Definitions of Morphological Parameters} \label{s_data_paramdef}

\begin{table*}
\centering
\begin{minipage}{101.5mm}
\renewcommand{\thefootnote}{\alph{footnote}}
\caption{Elliptical Measurement Regions Used for DDO~53 and NGC~6822}
\label{t_regspecial}
\begin{tabular}{@{}lccccc@{}}
\hline
    Name &
    \multicolumn{2}{c}{Centre (J2000)} &
    Major Axis &     Minor Axis &     Position \\
    &
    R.A.\footnotemark[1] &      Dec.\footnotemark[1] &
    (arcmin) &       (arcmin) &       Angle\footnotemark[2]\\
\hline
DDO~53 &     08 34 07.2 &   +66 10 54 &    2.0 &    2.0 &    $0\degr$ \\
NGC~6822 &   19 44 56.6 &   -14 47 21 &    27.0 &   15.0 &   $160\degr$ \\
\hline
\end{tabular}
$^a$ These are the centres of the optical discs given in RC3.\\
$^b$ The position angle is measured in degrees from north through east.
\end{minipage}
\end{table*}
\renewcommand{\thefootnote}{\arabic{footnote}}

Five quantities are used to define the infrared morphologies of these
galaxies.

The first parameter used is the concentration parameter.  The
parameter is given by \citet{bjc00}, \citet{c03}, and \citet{lpm04} as
\begin{equation}
C=5 \log \left(\frac{r_{80}}{r_{20}}\right)
\end{equation}
where $r_{80}$ and $r_{20}$ are the radii of the circles that contain
80\% and 20\% of the total light, respectively.  The total light from
these galaxies is measured within the optical disks defined by RC3;
more details are given later in this section.  The centres of the
circles for these measurements were chosen to match the centres of the
galaxies given in RC3.  The uncertainties for $C$ are calculated
assuming uncertainties of 1 pixel (1.5~arcsec) in the individual
radii, which also effectively accounts for uncertainties related to
the correspondence between the centre positions given by RC3 and the
true centres of the galaxies in these images.  This parameter is
generally used to indicate the central concentration of the light,
although the form of the parameter used here is really a measure of
the concentration of the central 20\% of the light.  A high value of
$C$ indicates that the emission originates primarily from near the
centre, whereas a low value indicates that most of the emission is
extended.

\citet{lpm04} also suggested using the normalized second order moment
of the brightest 20\% of the emission to indicate the central
concentration of light.  The unnormalized second order moment of the
light is given as
\begin{equation}
M=\sum_{i,j} f(i,j) r(i,j)^2
\end{equation}
where $f(i,j)$ is the value of pixel $i,j$ and $r(i,j)$ is the distance from
the centre of the target (which is set to the position given in RC3)
to pixel $i,j$.  The normalized second order moment of the brightest
20\% of the emission is given by
\begin{equation}
\overline{M}_{20}=\log\left(\frac{M_{20}}{M_{tot}}\right)
\end{equation}
where $M_{20}$ is the moment of the pixels that constitute the
brightest 20\% of the emission and $M_{tot}$ is the moment of all the
pixels.  (Note that the symbols used here are slightly different than
what is used in \citet{lpm04}.)  This value is always negative; lower
(more negative) values indicate that the emission is more centralized,
and higher (less negative) values indicate that the emission is
extended.  \citet{lpm04} suggested using $\overline{M}_{20}$ as an
alternative to the central concentration parameter $C$ used by
\citet{bjc00} and \citet{c03} because $\overline{M}_{20}$ is weighted
towards the most luminous regions.  The $\overline{M}_{20}$ parameter
measured for some sources (mainly point-like sources) may change by
$\sim5$\% or more if the in the central position is altered by 1 pixel
(1.5~arcsec), although the variations are relatively small for most of
the galaxies in the sample.  Nonetheless, we used a rigorous process
described in Appendix~\ref{s_findcentre} to find the central position
used for measuring $\overline{M}_{20}$.

Also of interest is how the infrared light is concentrated compared to
the optical light.  To do this, we define the normalized effective
radius of the light to be
\begin{equation}
\overline{R}_{eff}=\frac{R_{IR~eff}}{R_{opt}}
\end{equation}.
$R_{IR~eff}$ is the semi-major axis of the ellipse (with the same
centre, orientation, and ellipticity of the optical disc given in RC3)
that contains half of the light in the individual infrared wave band
that was measured within the optical disk defined by RC3.  $R_{opt}$
is the optical semi-major axis from RC3.  This parameter indicates how
the dust is distributed relative to the stars in the galaxy.  High
$\overline{R}_{eff}$ values indicate that the infrared emission is
extended compared to the starlight; low $\overline{R}_{eff}$ values
indicate that the infrared emission originates primarily from the
nuclei of the galaxies.  We use the logarithm of $\overline{R}_{eff}$ in
this analysis.  As for $C$, the uncertainties for $\overline{R}_{eff}$
are calculated assuming an uncertainties of 1 pixel (1.5~arcsec) in
$R_{IR~eff}$, which also accounts for uncertainties in the centre
positions of the galaxies.

The asymmetry in an image can be measured by subtracting an image
rotated $180^\circ$ from the original image.  This is given by
\citet{atsegv96} and \citet{c03} as
\begin{equation}
A=\frac{\sum_{i,j} |f(i,j)-f_{180}(i,j)|}{\sum_{i,j} |f(i,j)|}
\end{equation}
where $f(i,j)$ is the value of pixel $i,j$ and $f_{180}(i,j)$ is the
value of pixel $i,j$ in the image that has been rotated $180^\circ$
around the centre of the galaxy.  Higher values of $A$ indicate that
more emission originates from asymmetric structures.  Unlike the other
parameters used in this analysis, the asymmetry parameter may be
highly sensitive to errors in the centre of rotation.  We describe the
process used to identify the centre or rotation in
Appendix~\ref{s_findcentre}.

We use the Gini coefficient as defined in \citet{lpm04} to
determine the smoothness or peakedness of the image.  (The smoothing
parameter $S$ defined by \citet{c03} relies too heavily
on arbitrary choices concerning a smoothing width.  Moreover, it is
sensitive to the diameter of the galaxy.) To calculate the Gini
coefficient, the sky-subtracted pixels must first be ordered from the
lowest absolute pixel value to the highest absolute pixel value.  The
Gini coefficient is then calculated using
\begin{equation}
G=\frac{1}{\overline{|f|}n(n-1)}\sum_{i=1}^n(2i-n-1)|f(i)|
\end{equation}
where $\overline{|f|}$ is the average absolute pixel value, $n$
is the number of pixels, and $f(i)$ is the value of pixel $i$ after the
pixels are reordered by brightness.  A high value of $G$ indicates that
emission is concentrated in one or a few peaks, whereas a low value of
$G$ indicates smooth emission.

The parameters are measured in both the 24~$\mu$m data in its native
6 arcsec resolution and in the 3.6~$\mu$m data with PSFs
that are matched to the PSFs of the 24~$\mu$m data.  This is done by
convolving the 3.6~$\mu$m images with the convolution kernels of
K. D. Gordon, which not only degrade the resolution of the data but
also match the PSF of the 3.6~$\mu$m data to the PSF of the 24~$\mu$m
data\footnote[13]{\raggedright The kernels are available at
http://dirty.as.arizona.edu/$\sim$kgordon/mips/conv\_psfs/conv\_psfs.html.}.
The convolved 3.6~$\mu$m data were also rebinned into
1.5 arcsec square pixels to match the pixel scale of the
24~$\mu$m data, and the measurements are performed in the same regions
used for the 24~$\mu$m data.  The convolution and rebinning of the
3.6~$\mu$m data ensures that resolution and pixelation effects do not
create biases when the images are compared to the 24~$\mu$m images.

In both the 3.6 and 24~$\mu$m data, the parameters are measured within
the optical discs of the galaxies as defined in RC3.  Note that, if
RC3 does not list the ratio of the ellipse's axes or the position
angle of the major axis, the measurements are made within a circular
region with a diameter equal to the optical diameter listed in RC3.
This is only the case for a few galaxies that appear to be close to
circular (such as the face-on spiral galaxy NGC~3938) or that have
ill-defined shapes (such as Ho~I).  In many of the 3.6~$\mu$m and most
of the 24~$\mu$m images used in this paper, the surface brightness
drops below $3\sigma$ detection levels near or inside the optical
discs of the galaxies, and foreground stars become the dominant source
of emission outside the optical disc in the 3.6~$\mu$m images, so
inclusion of emission outside the optical disc will only add noise.
Exceptions are made for DDO~53 and NGC~6822, where the dust emission
clearly extends beyond the galaxies' optical discs.  In these cases,
the measurements were made in the elliptical regions that are listed
in Table~\ref{t_regspecial}, which are larger than the optical discs
given in RC3.

Note that, while we use of the optical disks defined in RC3 as the
regions in which we measure the morphological parameters, \citet{c03}
and \citet{lpm04} define the measurement regions using the Petrosian
radius.  We used the RC3 optical disks instead mainly because we
wanted to include the entire area within the optical disks of the
sample galaxies when measuring these parameters at 24~$\mu$m, whereas
areas based on the Petrosian radius may exclude part of the optical
disks of some galaxies with 24~$\mu$m emission.  Moreover, using the
optical disks allowed for measuring the same regions in both the 3.6
and 24~$\mu$m optical disks. Since the spatial extent of the 3.6 and
24~$\mu$m emission may vary within individual galaxies, the Petrosian
radii and measurement regions based on the Petrosian radii may vary as
well.  Furthermore, the galaxies in our sample are all located nearby,
so the problems described by \citet{cbj00} in defining optical disks
and in accounting for redshift and evolution effects (which were part
of the justification for using regions based on Petrosian radii) are
not as important for our analysis.

Additional sources that fall within the measurement regions (such as
foreground stars and companion galaxies) will make the galaxies appear
unusually extended and asymmetric in this analysis.  Therefore, these
sources need to be masked out before measurements are made.  Bright
foreground stars, identified by eye as unresolved sources with
relatively high (generally $\gtrsim5$) 3.6/24~$\mu$m flux density
ratios, were masked out of the 3.6~$\mu$m data and, in a few cases,
the 24~$\mu$m data.  Additionally, a 2~arcmin diameter circular
region at the centre of NGC~5195 was excluded from the optical disc of
NGC~5194, and NGC~1097A was masked out in the image of NGC~1097.

\begin{table}
\begin{center}
\caption{Morphological Parameters Measured in a Model Image of the 
    24~$\mu$m PSF \label{t_param_psf}}
\begin{tabular}{@{}cc@{}}
\hline
    Parameter &
    Value \\
\hline
$C$ &                     3.22 \\
$\overline{M}_{20}$ &     -2.72 \\
$A$ &                     0.080 \\
$G$ &                     0.94 \\
\hline
\end{tabular}
\end{center}
\end{table}

For reference, the $C$, $\overline{M}_{20}$, $A$, and $G$
morphological parameters were measured within a 4~arcmin diameter
circular region in a model image of the 24~$\mu$m PSF created with
STinyTim \footnote[14]{\raggedright Available from
http://ssc.spitzer.caltech.edu/archanaly/contributed/browse.html.}, a
PSF simulator designed for {\it Spitzer} \citep{k02}.  The parameters
are listed in Table~\ref{t_param_psf}.  Note that the
$\overline{M}_{20}$ and $G$ parameters may vary by $\sim10$\% if the
aperture is increased to 6~arcmin or decreased to 2~arcmin.  The
$C$ and $A$ parameters, however, do not vary significantly with the
diameter of the measurement region.  Also note that, in observations
of completely unresolved sources, much of the extended structure of
the PSF would not be detectable above the background noise, which
could change some of the measured parameters, particularly the
$\overline{M}_{20}$ and $G$ parameters.

\section{Robustness Tests of the Morphological Parameters} \label{s_test}

Before using the morphological parameters of the 24~$\mu$m data to
examine morphological trends within the sample, we performed a couple
of tests.  We first determine how the morphological parameters depend
on distance.  Next, we determine how the morphological parameters
depend on galaxy inclination.

\subsection{Dependence of Morphological Parameters on Distance}
    \label{s_test_distance}

\renewcommand{\thefootnote}{\alph{footnote}}
\begin{table*}
\centering
\begin{minipage}{149mm}
\renewcommand{\thefootnote}{\alph{footnote}}
\caption{Comparison of 3.6~$\mu$m Morphological Parameters Measured in 
    Original Images and Images Simulated to Represent 30~Mpc Distances 
    \label{t_paramirac1_dist}}
\begin{tabular}{@{}lcccccc@{}}
\hline
    Name &
    Distance (Mpc)\footnotemark[1] &
    $C$ &
    $\overline{M}_{20}$ &
    $\log(\overline{R}_{eff})$ &
    $A$ &
    $G$ \\
\hline
IC~2574 &
    3.5 (orig)&
    $2.57 \pm 0.02$ &
    $-0.888 \pm 0.008$ &
    $-0.518 \pm 0.003$ &
    $0.6885 \pm 0.0002$ &
    0.44 \\
    & 30 (sim) &
    $2.82 \pm 0.03$ &
    $-1.170 \pm 0.013$ &
    $-0.510 \pm 0.003$ &
    $0.54 \pm 0.03$ &
    0.34\\
NGC~2403 &
    3.5 (orig)&
    $3.38 \pm 0.03$ &
    $-1.198 \pm 0.006$ &
    $-0.781 \pm 0.003$ &
    $0.4510 \pm 0.0015$ &
    0.71 \\
    & 30 (sim) &
    $3.24 \pm 0.03$ &
    $-1.357 \pm 0.007$ &
    $-0.754 \pm 0.003$ &
    $0.313 \pm 0.012$ &
    0.67\\
NGC~3031 &
    3.5 (orig)&
    $4.18 \pm 0.04$ &
    $-2.37 \pm 0.04$ &
    $-0.932 \pm 0.003$ &
    $0.103 \pm 0.004$ &
    0.75 \\
    & 30 (sim) &
    $3.83 \pm 0.04$ &
    $-2.42 \pm 0.03$ &
    $-0.895 \pm 0.003$ &
    $0.10 \pm 0.07$ &
    0.73 \\
NGC~4236 &
    3.5 (orig)&
    $3.043 \pm 0.017$ &
    $-0.822 \pm 0.007$ &
    $-0.5938 \pm 0.0019$ &
    $0.51575 \pm 0.00012$ &
    0.38 \\
    & 30 (sim) &
    $3.46 \pm 0.02$ &
    $-1.69 \pm 0.03$ &
    $-0.5940 \pm 0.0019$ &
    $0.345 \pm 0.014$ &
    0.38\\
NGC~4736 &
    5.3 (orig)&
    $4.48 \pm 0.14$ &
    $-2.73 \pm 0.03$ &
    $-1.019 \pm 0.010$ &
    $0.06 \pm 0.04$ &
    0.87 \\
    & 30 (sim) &
    $3.12 \pm 0.08$ &
    $-2.25 \pm 0.04$ &
    $-0.945 \pm 0.009$ &
    $0.09 \pm 0.10$ &
    0.86\\
NGC~6946 &
    5.5 (orig)&
    $2.85 \pm 0.03$ &
    $-1.194 \pm 0.011$ &
    $-0.390 \pm 0.002$ &
    $0.3956 \pm 0.0005$ &
    0.58 \\
    & 30 (sim) &
    $2.80 \pm 0.03$ &
    $-1.130 \pm 0.008$ &
    $-0.369 \pm 0.002$ &
    $0.343 \pm 0.007$ &
    0.55\\
\hline
\end{tabular}
$^a$ The measurements in the original images have distances labelled as
    (orig).  The measurements in the simulated images have distances
    labelled as (sim).
\end{minipage}
\end{table*}

\begin{table*}
\centering
\begin{minipage}{144.5mm}
\renewcommand{\thefootnote}{\alph{footnote}}
\caption{Comparison of 24~$\mu$m Morphological Parameters Measured in 
    Original Images and Images Simulated to Represent 30~Mpc Distances 
    \label{t_param24_dist}}
\begin{tabular}{@{}lcccccc@{}}
\hline
    Name &
    Distance (Mpc)\footnotemark[1] &
    $C$ &
    $\overline{M}_{20}$ &
    $\log(\overline{R}_{eff})$ &
    $A$ &
    $G$\\
\hline
IC~2574 &
    3.5 (orig)&
    $1.582 \pm 0.014$ &
    $-0.653 \pm 0.016$ &
    $-0.3259 \pm 0.0017$ &
    $1.884 \pm 0.005$ &
    0.62 \\
    & 30 (sim) &
    $1.99 \pm 0.02$ &
    $-0.6 \pm 0.4$ &
    $-0.3418 \pm 0.0018$ &
    $1.1 \pm 0.4$ &
    0.54 \\
NGC~2403 &
    3.5 (orig)&
    $2.53 \pm 0.03$ &
    $-1.138 \pm 0.015$ &
    $-0.925 \pm 0.004$ &
    $1.009 \pm 0.0004$ &
    0.82 \\
    & 30 (sim) &
    $2.64 \pm 0.03$ &
    $-1.37 \pm 0.014$ &
    $-0.863 \pm 0.004$ &
    $0.59 \pm 0.02$ &
    0.80 \\
NGC~3031 &
    3.5 (orig)&
    $2.95 \pm 0.02$ &
    $-1.01 \pm 0.02$ &
    $-0.6818 \pm 0.0019$ &
    $0.763 \pm 0.005$ &
    0.71 \\
    & 30 (sim) &
    $2.95 \pm 0.03$ &
    $-1.3 \pm 0.3$ &
    $-0.6676 \pm 0.0019$ &
    $0.38 \pm 0.04$ &
    0.69 \\
NGC~4236 &
    3.5 (orig)&
    $1.414 \pm 0.013$ &
    $-0.896 \pm 0.014$ &
    $-0.600 \pm 0.002$ &
    $1.786 \pm 0.003$ &
    0.67 \\
    & 30 (sim) &
    $1.858 \pm 0.016$ &
    $-0.8 \pm 0.5$ &
    $-0.5618 \pm 0.0018$ &
    $0.95 \pm 0.16$ &
    0.66 \\
NGC~4736 &
    5.3 (orig)&
    $2.96 \pm 0.10$ &
    $-1.661 \pm 0.017$ &
    $-0.971 \pm 0.009$ &
    $0.30 \pm 0.02$ &
    0.89 \\
    & 30 (sim) &
    $2.53 \pm 0.08$ &
    $-2.07 \pm 0.06$ &
    $-0.899 \pm 0.008$ &
    $0.15 \pm 0.02$ &
    0.66 \\
NGC~6946 &
    5.5 (orig)&
    $5.82 \pm 0.13$ &
    $-2.02 \pm 0.11$ &
    $-0.525 \pm 0.003$ &
    $0.656 \pm 0.014$ &
    0.78 \\
    & 30 (sim) &
    $4.42 \pm 0.07$ &
    $-2.47 \pm 0.03$ &
    $-0.510 \pm 0.003$ &
    $0.453 \pm 0.019$ &
    0.74 \\
\hline
\end{tabular}
$^a$ The measurements in the original images have distances labelled as
    (orig).  The measurements in the simulated images have distances
    labelled as (sim).
\end{minipage}
\end{table*}
\renewcommand{\thefootnote}{\arabic{footnote}}

The galaxies in this sample are not distributed homogeneously with
respect to distance.  The more distant galaxies tend to be early-type
spiral and elliptical galaxies, whereas the nearby galaxies tend to be
late-type spiral and irregular galaxies.  Moreover, the distant
galaxies tend to be relatively unusual objects that were chosen to
completely sample the colour-morphology-luminosity space of nearby
galaxies.  Therefore, we must test how the morphological parameters
vary with distance so as to disentangle distance-related effects from
true variations in the morphologies related to Hubble-type.

To perform this test, we used the 3.6 and 24~$\mu$m images of all
galaxies in the sample within 6~Mpc that have optical diameters of at
least 10~arcmin except for NGC~6822.  This subsample contains 6
galaxies ranging from Sab to Sm.  We smoothed and rebinned the images
to simulate what the galaxies would look like at a distance of 30~Mpc
(approximately the maximum distance for the galaxies in the SINGS
sample) in images with the same resolution and pixel scale of the
24~$\mu$m images.  At a distance of 30~Mpc, all galaxies in the test
subsample are still larger than 1~arcmin and are therefore comparable
to the most distant galaxies in the sample.  (NGC~6822, which is the
closest galaxy in the SINGS sample, would have an angular size of
$\sim30$~arcsec at a distance of 30~Mpc.  This is smaller than any of
the galaxies in this paper's sample and also poorly resolved at the
resolution of the 24~$\mu$m data.)

\begin{figure}
\begin{center}
\epsfig{file=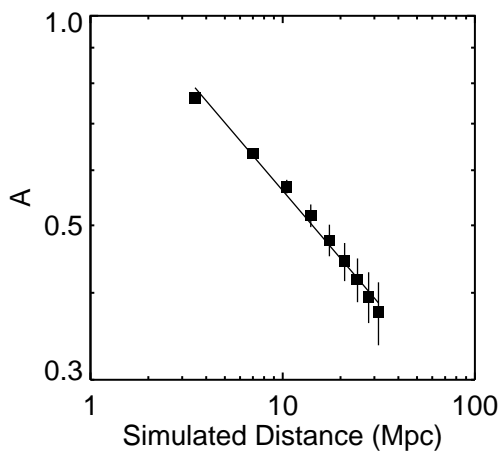, height=67mm}
\end{center}
\caption{Variations in the $A$ parameter measured in images of
NGC~3031 simulated to represent the galaxy's appearance at distances
between 3.5 and $\sim30$~Mpc (see Section~\ref{s_test_distance}).  The
line represents the best-fitting power law, which has a slope of
-0.30.}
\label{f_example_acorr}
\end{figure}

A comparison between the morphological parameters measured in these
smoothed images and the parameters measured in the original images is
presented in Tables~\ref{t_paramirac1_dist} and \ref{t_param24_dist}.
From these numbers, it can be seen that the $C$,
$\log(\overline{R}_{eff})$, and $G$ are all relatively invariant
with distance in both wave bands.  The $\overline{M}_{20}$
parameter only varies $\sim20$\% in both wave bands for most galaxies,
although it does increase by a factor of $\sim2$ for the 3.6~$\mu$m
image of NGC~4236; we will simply note these limitations in the
analysis rather than develop a correction function.  The 3.6~$\mu$m $A$
parameter does not strongly vary except for the irregular galaxies
IC~2574 and NGC~4236.  Again, we will simply note that the 3.6~$\mu$m
$A$ parameter for the irregular galaxies may be highly uncertain, but we
will skip applying a correction factor for this analysis. The
24~$\mu$m $A$ parameter, however, decreases by approximately a factor of
2 for all galaxies.  This indicates that the 24~$\mu$m $A$ parameter is
dependent on distance for all galaxies and that a correction needs to
be applied.

\begin{table*}
\centering
\begin{minipage}{133mm}
\renewcommand{\thefootnote}{\alph{footnote}}
\caption{Comparison of 3.6~$\mu$m Morphological Parameters Measured in
    Original Images and Images Simulated to Various Inclinations}
\label{t_paramirac1_inc}
\begin{tabular}{@{}lcccccc@{}}
\hline
    Name &
    Inclination\footnotemark[1] &
    $C$ &
    $\overline{M}_{20}$ &
    $\log(\overline{R}_{eff})$ &
    $A$ &
    $G$\\
\hline
NGC~628 &
    $0\degr$ (orig)&
    $2.97 \pm 0.04$ &
    $-1.592 \pm 0.017$ &
    $-0.482 \pm 0.017$ &
    $0.273 \pm 0.004$ &
    0.65\\
    & $40\degr$ (sim)&
    $2.97 \pm 0.04$ &
    $-1.574 \pm 0.019$ &
    $-0.60 \pm 0.02$ &
    $0.2736 \pm 0.0015$ &
    0.65\\
    & $80\degr$ (sim)&
    $3.78 \pm 0.07$ &
    $-1.520 \pm 0.011$ &
    $-1.04 \pm 0.06$ &
    $0.2750 \pm 0.0005$ &
    0.64\\
NGC~3184 &
    $0\degr$ (orig)&
    $2.43 \pm 0.04$ &
    $-1.63 \pm 0.02$ &
    $-0.386 \pm 0.019$ &
    $0.241 \pm 0.002$ &
    0.55\\
    & $40\degr$ (sim)&
    $2.48 \pm 0.04$ &
    $-1.68 \pm 0.02$ &
    $-0.50 \pm 0.02$ &
    $0.241 \pm 0.002$ &
    0.55\\
    & $80\degr$ (sim)&
    $3.43 \pm 0.08$ &
    $-1.94 \pm 0.03$ &
    $-0.94 \pm 0.07$ &
    $0.2419 \pm 0.0006$ &
    0.55\\
NGC~3938 &
    $0\degr$ (orig)&
    $2.96 \pm 0.07$ &
    $-2.03 \pm 0.03$ &
    $-0.49 \pm 0.03$ &
    $0.214 \pm 0.004$ &
    0.63\\
    & $40\degr$ (sim)&
    $2.91 \pm 0.08$ &
    $-2.03 \pm 0.03$ &
    $-0.60 \pm 0.04$ &
    $0.214 \pm 0.004$ &
    0.63\\
    & $80\degr$ (sim)&
    $3.63 \pm 0.14$ &
    $-2.09 \pm 0.03$ &
    $-1.03 \pm 0.12$ &
    $0.230 \pm 0.004$ &
    0.63\\
\hline
\end{tabular}
$^a$ The measurements in the original images have inclinations
     labelled as $0\degr$ (orig).  The measurements in the simulated
     images have inclinations labelled as (sim).
\end{minipage}
\end{table*}

\begin{table*}
\centering
\begin{minipage}{133mm}
\renewcommand{\thefootnote}{\alph{footnote}}
\caption{Comparison of 24~$\mu$m Morphological Parameters Measured in 
    Original Images and Images Simulated to Various Inclinations}
\label{t_param24_inc}
\begin{tabular}{@{}lcccccc@{}}
\hline
    Name &
    Inclination\footnotemark[1] &
    $C$ &
    $\overline{M}_{20}$ &
    $\log(\overline{R}_{eff})$ &
    $A$ &
    $G$ \\
\hline
NGC~628 &
    $0\degr$ (orig)&
    $1.94 \pm 0.03$ &
    $-0.845 \pm 0.014$ &
    $-0.446 \pm 0.015$ &
    $0.9016 \pm 0.0007$ &
    0.73\\
    & $40\degr$ (sim)&
    $1.96 \pm 0.03$ &
    $-0.884 \pm 0.015$ &
    $-0.58 \pm 0.02$ &
    $0.9002 \pm 0.0010$ &
    0.73\\
    & $80\degr$ (sim)&
    $2.76 \pm 0.05$ &
    $-1.02 \pm 0.02$ &
    $-1.01 \pm 0.06$ &
    $0.841 \pm 0.009$ &
    0.74\\
NGC~3184 &
    $0\degr$ (orig)&
    $2.02 \pm 0.04$ &
    $-0.997 \pm 0.018$ &
    $-0.40 \pm 0.02$ &
    $0.728 \pm 0.006$ &
    0.68\\
    & $40\degr$ (sim)&
    $2.16 \pm 0.04$ &
    $-1.021 \pm 0.019$ &
    $-0.51 \pm 0.03$ &
    $0.723 \pm 0.010$ &
    0.68\\
    & $80\degr$ (sim)&
    $3.44 \pm 0.09$ &
    $-1.09 \pm 0.06$ &
    $-0.94 \pm 0.07$ &
    $0.688 \pm 0.010$ &
    0.68\\
NGC~3938 &
    $0\degr$ (orig)&
    $2.37 \pm 0.06$ &
    $-0.858 \pm 0.008$ &
    $-0.47 \pm 0.03$ &
    $0.6574 \pm 0.0006$ &
    0.71\\
    & $40\degr$ (sim)&
    $2.39 \pm 0.06$ &
    $-0.817 \pm 0.008$ &
    $-0.59 \pm 0.04$ &
    $0.6553 \pm 0.0009$ &
    0.71\\
    & $80\degr$ (sim)&
    $3.22 \pm 0.11$ &
    $-0.694 \pm 0.007$ &
    $-0.97 \pm 0.10$ &
    $0.612 \pm 0.008$ &
    0.71\\
\hline
\end{tabular}
$^a$ The measurements in the original images have inclinations labelled
     as $0\degr$ (orig).  The measurements in the simulated images have
     inclinations labelled as (sim).
\end{minipage}
\end{table*}
\renewcommand{\thefootnote}{\arabic{footnote}}

To find the correction for the 24~$\mu$m $A$ parameter, we produced
simulated images of each test subsample galaxies at several artificial
distances between their actual distance and 30~Mpc and measured the $A$
parameter.  We empirically examined the data and determined that the
function that best describes the relation between $A$ and distance is a
power law.  We used a least-squares fit to the data to find the index
of the power law that describes $A$ as a function of simulated distance
for each galaxy.  An example using NGC~3031 is shown in
Figure~\ref{f_example_acorr}.  Afterwards, we averaged the power law
indices from the fits to all galaxies.  The result is
\begin{equation}
A \propto d^{-0.26\pm0.03}
\end{equation}
To correct the 24~$\mu$m $A$ parameters, we will adjust all measured $A$
values to their equivalent at a distance of 10~Mpc (the approximate
median distance of the galaxies in this sample) using the equation
\begin{equation}
A_{corrected}=0.55 A_{observed}  d^{0.30}
\label{e_acorr_dist}
\end{equation}
where $d$ is in Mpc.  This equation may only be applicable to the
24~$\mu$m data, as the power law probably varies with wave band and
resolution.

This analysis does indicate that the $A$ and $\overline{M}_{20}$
parameters may be sensitive to distance in other wave bands that trace
either dust or star formation.  Therefore, their usefulness at
redshifts higher than those in this sample may be questionable.  As is
also discussed by \citep{lpm04}, simulated high-redshift images of
nearby galaxies are needed to compare to observed high-redshift
galaxies.  The distance-sensitivity of the two parameters also implies
that they are dependent on the angular resolution at which the
galaxies are observed.  Therefore, to accurately compare parameters
measured within images of two different wave bands, the images must
have matching resolutions.

\subsection{Dependence of Morphological Parameters on Inclination}
    \label{s_test_inclination}

\begin{figure}
\begin{center}
\epsfig{file=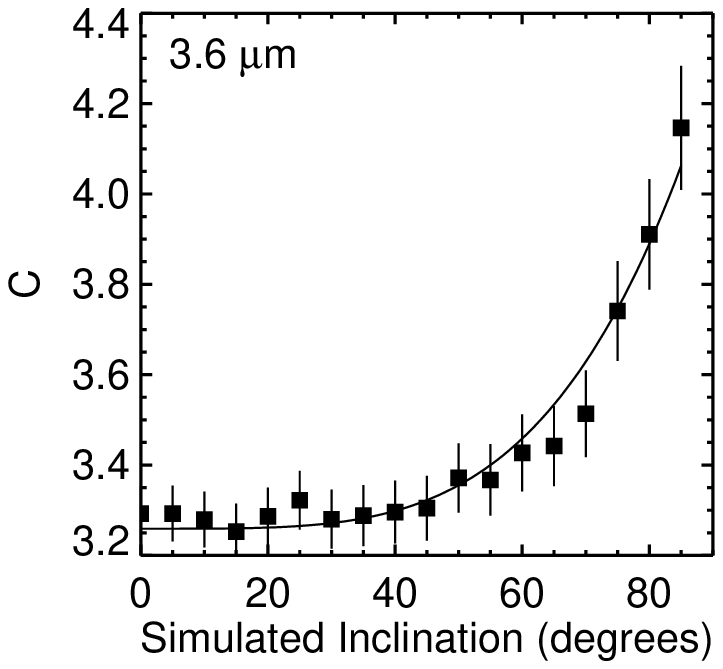, height=67mm}
\epsfig{file=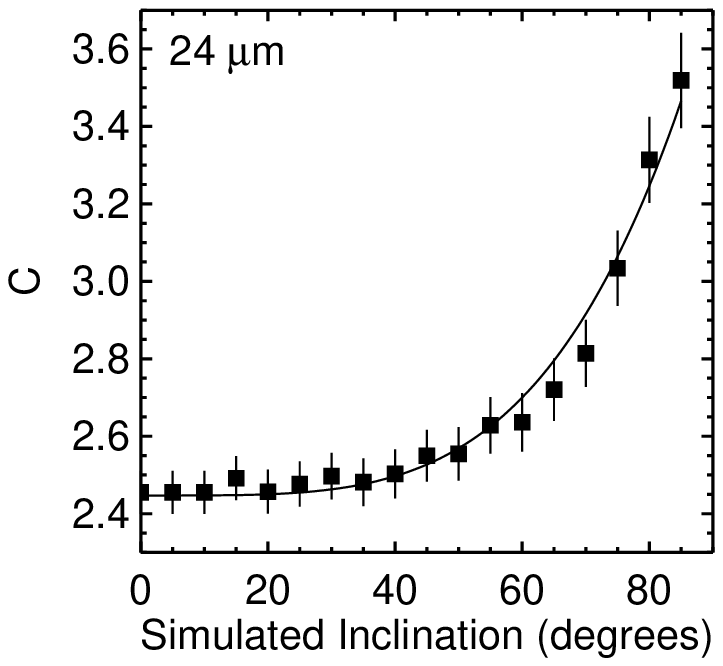, height=67mm}
\end{center}
\caption{Variations in the $C$ parameter measured in images of
NGC~3938 simulated to represent the galaxy's appearance at various
inclinations (see Section~\ref{s_test_inclination}).  The lines represent
the best-fitting function.}
\label{f_example_ccorr}
\end{figure}

\begin{figure}
\begin{center}
\epsfig{file=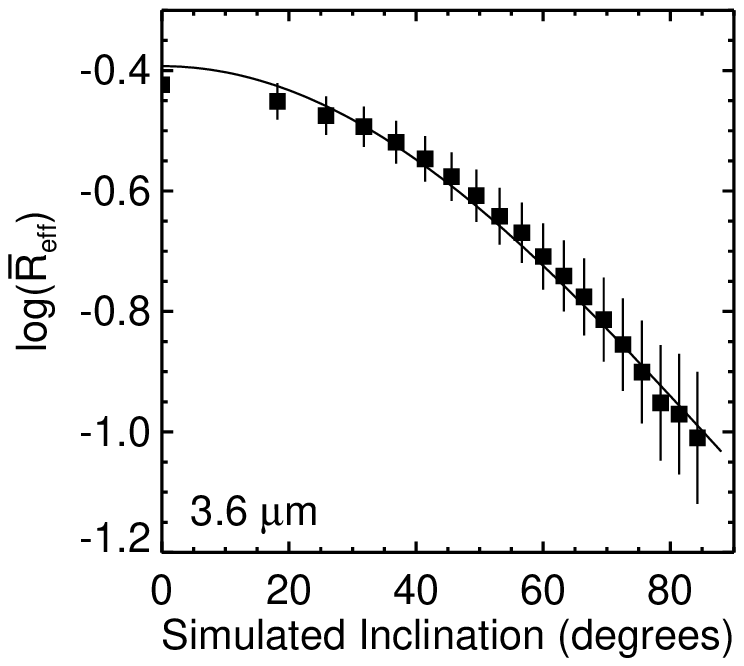, height=67mm}
\epsfig{file=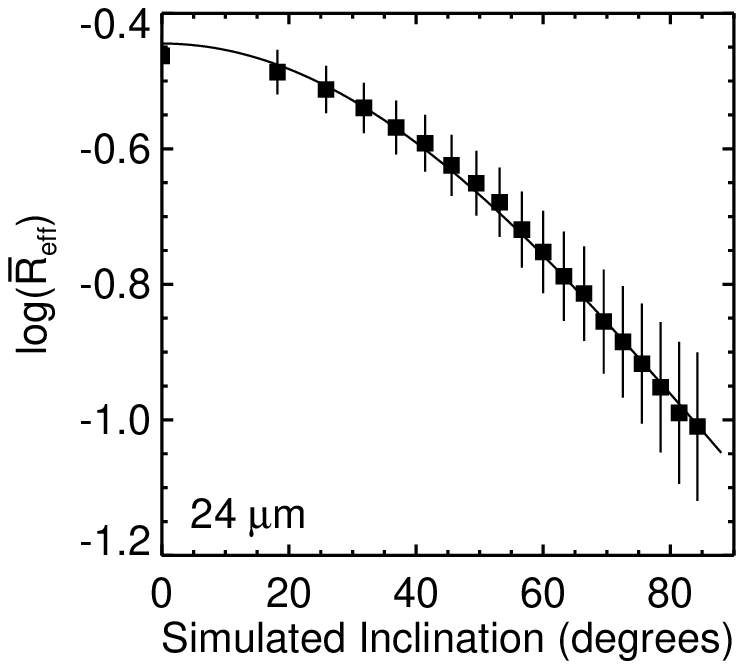, height=67mm}
\end{center}
\caption{Variations in the $\log(\overline{R}_{eff})$ parameter
measured in images of NGC~3938 simulated to represent the galaxy's
appearance at various inclinations (see
Section~\ref{s_test_inclination}).  The lines represent the best-fitting
function.}
\label{f_example_lreffcorr}
\end{figure}

Since the galaxies in this sample exhibit a wide range of inclinations
and since galaxies' appearance change when viewed edge-on, we
simulated the effects of inclining a galaxy between $0\degr$ and
$80\degr$ using NGC~628, NGC~3184, and NGC~3938, which have
minor/major axis ratios of 0.9 or higher and optical diameters greater
than 5~arcmin.  In this analysis, these galaxies are treated as
having inclinations of $0\degr$, although the axis ratios may imply a
slight inclination.  The simulated inclinations were performed simply
by projecting the images onto a plane tilted to the desired angle.
For this analysis, the emission is assumed to be infinitely thin, 
although we note that significant stellar emission may extend outside the
plane of S0 and early-type spiral galaxies.

A comparison between the 3.6 and 24~$\mu$m morphological parameters
measured in the original images and images with simulated inclinations
of $40\degr$ and $80\degr$ is shown in Table~\ref{t_paramirac1_inc}
and \ref{t_param24_inc}.  For this section alone, the measurements
region is treated as circular in the original images.  In the images
with simulated inclinations, the ratio of the axes of the measurement
region is set to the cosine of the inclination angle.  The $A$,
$\overline{M}_{20}$, and $G$ parameters all vary less than
$\sim20$\% between the original and simulated images, but the
comparison demonstrates that $C$ and $\log(\overline{R}_{eff})$
for both wave bands are sensitive to inclination effects.  These will
need to be corrected to deal with these inclination effects.

We derive relations for the corrections needed to be applied to these
parameters by measuring the $C$ and $\log(\overline{R}_{eff})$
parameters measured in the 3.6 and 24~$\mu$m images simulated with
inclinations between $0\degr$ and $85\degr$.  For the $C$ parameter, we
empirically determined that it varies as a function of a constant
added to $\theta^4$, where $\theta$ is the inclination angle in
degrees.  The $\log(\overline{R}_{eff})$ parameter was found to
vary as a function of $cos(\theta)$.  These functions were fit to $C$
and $\log(\overline{R}_{eff})$ for each galaxy in each wave
band, and then the median and uncertainty for the resulting
coefficients were calculated to give the following correction
functions:
\begin{equation}
C_{corrected}=C_{original}-(2.0 \pm 0.3)\times10^{-8}\theta^4
\label{e_ccorr_inc}
\end{equation}
\begin{eqnarray}
\log(\overline{R}_{eff})_{corrected} = \log(\overline{R}_{eff})_{original}
    \nonumber\\
    +(0.675\pm0.012)(1-\cos(\theta))
\label{e_lreffcorr_inc}
\end{eqnarray}
The values for $C_{corrected}$ and
$\log(\overline{R}_{eff})_{corrected}$ are the parameters that would
be measured if the galaxies were viewed face-on.  Example fits are
shown for NGC~3938 in Figures~\ref{f_example_ccorr} and
\ref{f_example_lreffcorr}.  Note that the variations in each wave band
are statistically similar.

The inclination-related corrections are only applied to disc galaxies
(S0-Sd galaxies).  To apply these corrections, the inclination
is calculated using
\begin{equation}
\theta=\cos^{-1}\left( \left( 
    \frac{(\frac{r_{minor}}{r_{major}})^2-q_o^2}{1-q_o^2} 
     \right)^{0.5} \right)
\end{equation} 
where $r_{minor}$ is the optical semiminor axis, $r_{major}$ is the
optical semimajor axis, and $q_o$ is the intrinsic optical axial
ratio, which is set to 0.20.  NGC~3190, NGC~4631, and NGC~5866 are
viewed close to edge-on, so the inclination is set to $90\degr$ for
these galaxies.  

Because the correction was calculated assuming that the stellar and
24~$\mu$m emission was infinitely thin (geometrically and optically),
the corrections may be inaccurate for the 3.6~$\mu$m parameters of
some nearly edge-on early-type spiral and S0 galaxies.  However, we
anticipate that the 24~$\mu$m emission should originate mostly from a
thin disc, so the corrections should be accurate for the 24~$\mu$m
parameters.  The 3.6~$\mu$m stellar emission from late-type galaxies
should also originate from a relatively thin disc, so the corrections
for the 3.6~$\mu$m emission for these galaxies should also be fairly
accurate.

\begin{table*}
\centering
\begin{minipage}{147.5mm}
\renewcommand{\thefootnote}{\alph{footnote}}
\caption{Morphological Parameters Measured in 3.6~$\mu$m 
    Data}
\label{t_paramirac1}
\begin{tabular}{@{}lcccccc@{}}
\hline
    Name &
    $C$\footnotemark[1] &
    $\overline{M}_{20}$ &
    $\overline{R}_{eff}$\footnotemark[2] &
    $\log(\overline{R}_{eff})$\footnotemark[2] &
    $A$&
    $G$\\
\hline
DDO~53 &
    $2.52 \pm 0.11$ &
    $-0.695 \pm 0.010$ &
    $0.726 \pm 0.016$ &
    $-0.139 \pm 0.010$ &
    $1.1360 \pm 0.0004$ &
    0.56 \\
Ho~I &
    $1.85 \pm 0.05$ &
    $-0.748 \pm 0.008$ &
    $0.592 \pm 0.007$ &
    $-0.228 \pm 0.005$ &
    $0.7670 \pm 0.0008$ &
    0.41 \\
Ho~II &
    $2.44 \pm 0.03$ &
    $-0.738 \pm 0.007$ &
    $0.415 \pm 0.003$ &
    $-0.381 \pm 0.003$ &
    $0.7823 \pm 0.0005$ &
    0.54 \\
IC~2574 &
    $2.57 \pm 0.02$ &
    $-0.888 \pm 0.008$ &
    $0.3034 \pm 0.0019$ &
    $-0.518 \pm 0.003$ &
    $0.6885 \pm 0.0002$ &
    0.44 \\
IC~3583 &
    $1.67 \pm 0.13$ &
    $-0.884 \pm 0.006$ &
    $0.251 \pm 0.011$ &
    $-0.60 \pm 0.02$ &
    $1.002 \pm 0.008$ &
    0.59 \\
IC~4710 &
    $2.07 \pm 0.06$ &
    $-0.883 \pm 0.004$ &
    $0.413 \pm 0.007$ &
    $-0.384 \pm 0.007$ &
    $0.512 \pm 0.002$ &
    0.41 \\
Mrk~33 &
    $2.8 \pm 0.4$ &
    $-2.02 \pm 0.03$ &
    $0.30 \pm 0.03$ &
    $-0.52 \pm 0.04$ &
    $0.10 \pm 0.09$ &
    0.67 \\
NGC~24 &
    $2.18 \pm 0.10$ &
    $-1.88 \pm 0.02$ &
    $0.711 \pm 0.004$ &
    $-0.148 \pm 0.003$ &
    $0.206 \pm 0.007$ &
    0.54 \\
NGC~337 &
    $2.28 \pm 0.13$ &
    $-1.649 \pm 0.017$ &
    $0.460 \pm 0.009$ &
    $-0.337 \pm 0.008$ &
    $0.2799 \pm 0.0010$ &
    0.63 \\
NGC~628 &
    $2.95 \pm 0.04$ &
    $-1.573 \pm 0.019$ &
    $0.358 \pm 0.002$ &
    $-0.446 \pm 0.003$ &
    $0.266 \pm 0.003$ &
    0.63 \\
NGC~855 &
    $3.2 \pm 0.2$ &
    $-2.136 \pm 0.019$ &
    $0.124 \pm 0.010$ &
    $-0.91 \pm 0.03$ &
    $0.10 \pm 0.02$ &
    0.59 \\
NGC~925 &
    $2.82 \pm 0.04$ &
    $-1.324 \pm 0.009$ &
    $0.501 \pm 0.002$ &
    $-0.301 \pm 0.002$ &
    $0.391 \pm 0.004$ &
    0.55 \\
NGC~1097 &
    $4.78 \pm 0.12$ &
    $-2.78 \pm 0.02$ &
    $0.264 \pm 0.003$ &
    $-0.579 \pm 0.004$ &
    $0.1932 \pm 0.0011$ &
    0.71 \\
NGC~1266 &
    $3.1 \pm 0.3$ &
    $-2.15 \pm 0.02$ &
    $0.545 \pm 0.016$ &
    $-0.263 \pm 0.013$ &
    $0.093 \pm 0.003$ &
    0.64 \\
NGC~1291 &
    $4.99 \pm 0.09$ &
    $-2.87 \pm 0.03$ &
    $0.282 \pm 0.003$ &
    $-0.549 \pm 0.004$ &
    $0.09 \pm 0.02$ &
    0.65 \\
NGC~1316 &
    $4.55 \pm 0.08$ &
    $-2.77 \pm 0.03$ &
    $0.245 \pm 0.002$ &
    $-0.612 \pm 0.004$ &
    $0.091 \pm 0.010$ &
    0.69 \\
NGC~1317 &
    $3.9 \pm 0.2$ &
    $-2.39 \pm 0.02$ &
    $0.313 \pm 0.009$ &
    $-0.504 \pm 0.013$ &
    $0.0628 \pm 0.0014$ &
    0.60 \\
NGC~1377 &
    $3.0 \pm 0.4$ &
    $-2.28 \pm 0.03$ &
    $0.321 \pm 0.014$ &
    $-0.494 \pm 0.019$ &
    $0.10 \pm 0.05$ &
    0.68 \\
NGC~1482 &
    $3.1 \pm 0.4$ &
    $-2.32 \pm 0.03$ &
    $0.209 \pm 0.010$ &
    $-0.68 \pm 0.02$ &
    $0.11 \pm 0.05$ &
    0.81 \\
NGC~1510 &
    $2.6 \pm 0.3$ &
    $-2.00 \pm 0.03$ &
    $0.438 \pm 0.019$ &
    $-0.358 \pm 0.019$ &
    $0.07 \pm 0.05$ &
    0.54 \\
NGC~1512 &
    $3.93 \pm 0.10$ &
    $-2.58 \pm 0.03$ &
    $0.313 \pm 0.003$ &
    $-0.504 \pm 0.004$ &
    $0.139 \pm 0.014$ &
    0.71 \\
NGC~1566 &
    $3.57 \pm 0.08$ &
    $-2.41 \pm 0.04$ &
    $0.273 \pm 0.003$ &
    $-0.564 \pm 0.005$ &
    $0.191 \pm 0.006$ &
    0.72 \\
NGC~1705 &
    $3.3 \pm 0.2$ &
    $-2.20 \pm 0.04$ &
    $0.360 \pm 0.013$ &
    $-0.444 \pm 0.016$ &
    $0.37 \pm 0.02$ &
    0.60 \\
NGC~2403 &
    $3.16 \pm 0.03$ &
    $-1.198 \pm 0.006$ &
    $0.3406 \pm 0.0011$ &
    $-0.4678 \pm 0.0015$ &
    $0.4510 \pm 0.0015$ &
    0.71 \\
NGC~2798 &
    $2.8 \pm 0.4$ &
    $-2.33 \pm 0.04$ &
    $0.248 \pm 0.010$ &
    $-0.606 \pm 0.017$ &
    $0.14 \pm 0.07$ &
    0.76 \\
NGC~2841 &
    $3.37 \pm 0.09$ &
    $-2.44 \pm 0.03$ &
    $0.354 \pm 0.003$ &
    $-0.451 \pm 0.004$ &
    $0.0618 \pm 0.0006$ &
    0.64 \\
NGC~2915 &
    $2.55 \pm 0.17$ &
    $-1.74 \pm 0.03$ &
    $0.262 \pm 0.013$ &
    $-0.58 \pm 0.02$ &
    $0.33 \pm 0.02$ &
    0.44 \\
NGC~2976 &
    $1.93 \pm 0.06$ &
    $-1.045 \pm 0.010$ &
    $0.586 \pm 0.004$ &
    $-0.232 \pm 0.003$ &
    $0.181 \pm 0.002$ &
    0.54 \\
NGC~3031 &
    $3.91 \pm 0.04$ &
    $-2.37 \pm 0.04$ &
    $0.2565 \pm 0.0009$ &
    $-0.5910 \pm 0.0016$ &
    $0.103 \pm 0.004$ &
    0.75 \\
NGC~3049 &
    $3.06 \pm 0.19$ &
    $-2.07 \pm 0.02$ &
    $0.498 \pm 0.011$ &
    $-0.303 \pm 0.010$ &
    $0.12 \pm 0.03$ &
    0.50 \\
NGC~3184 &
    $2.40 \pm 0.04$ &
    $-1.62 \pm 0.02$ &
    $0.440 \pm 0.003$ &
    $-0.357 \pm 0.003$ &
    $0.239 \pm 0.002$ &
    0.53 \\
NGC~3185 &
    $3.03 \pm 0.19$ &
    $-2.18 \pm 0.03$ &
    $0.471 \pm 0.011$ &
    $-0.327 \pm 0.010$ &
    $0.11 \pm 0.03$ &
    0.56 \\
NGC~3187 &
    $2.71 \pm 0.12$ &
    $-1.713 \pm 0.015$ &
    $0.904 \pm 0.008$ &
    $-0.044 \pm 0.004$ &
    $0.3399 \pm 0.0006$ &
    0.65 \\
NGC~3190 &
    $2.5 \pm 0.2$ &
    $-2.48 \pm 0.03$ &
    $0.434 \pm 0.006$ &
    $-0.363 \pm 0.006$ &
    $0.116 \pm 0.015$ &
    0.71 \\
NGC~3198 &
    $2.54 \pm 0.07$ &
    $-1.73 \pm 0.03$ &
    $0.352 \pm 0.003$ &
    $-0.453 \pm 0.004$ &
    $0.153 \pm 0.003$ &
    0.67 \\
NGC~3265 &
    $2.6 \pm 0.4$ &
    $-1.99 \pm 0.03$ &
    $0.194 \pm 0.019$ &
    $-0.71 \pm 0.04$ &
    $0.10 \pm 0.05$ &
    0.73 \\
NGC~3351 &
    $3.97 \pm 0.10$ &
    $-2.57 \pm 0.03$ &
    $0.349 \pm 0.003$ &
    $-0.457 \pm 0.004$ &
    $0.081 \pm 0.007$ &
    0.64 \\
NGC~3521 &
    $3.21 \pm 0.08$ &
    $-2.36 \pm 0.03$ &
    $0.259 \pm 0.002$ &
    $-0.586 \pm 0.004$ &
    $0.086 \pm 0.017$ &
    0.74 \\
NGC~3621 &
    $3.23 \pm 0.05$ &
    $-1.554 \pm 0.010$ &
    $0.339 \pm 0.002$ &
    $-0.470 \pm 0.003$ &
    $0.40076 \pm 0.00012$ &
    0.72 \\
NGC~3627 &
    $2.93 \pm 0.07$ &
    $-2.04 \pm 0.04$ &
    $0.365 \pm 0.003$ &
    $-0.438 \pm 0.003$ &
    $0.290 \pm 0.003$ &
    0.67 \\
NGC~3773 &
    $2.8 \pm 0.3$ &
    $-1.96 \pm 0.02$ &
    $0.38 \pm 0.02$ &
    $-0.42 \pm 0.02$ &
    $0.08 \pm 0.05$ &
    0.54 \\
NGC~3938 &
    $2.95 \pm 0.07$ &
    $-2.03 \pm 0.03$ &
    $0.376 \pm 0.005$ &
    $-0.425 \pm 0.005$ &
    $0.214 \pm 0.004$ &
    0.63 \\
NGC~4125 &
    $4.00 \pm 0.14$ &
    $-2.49 \pm 0.03$ &
    $0.148 \pm 0.004$ &
    $-0.831 \pm 0.013$ &
    $0.06 \pm 0.02$ &
    0.65 \\
NGC~4236 &
    $3.043 \pm 0.017$ &
    $-0.822 \pm 0.007$ &
    $0.2548 \pm 0.0011$ &
    $-0.5938 \pm 0.0019$ &
    $0.51575 \pm 0.00012$ &
    0.47 \\
NGC~4254 &
    $3.13 \pm 0.08$ &
    $-2.10 \pm 0.03$ &
    $0.367 \pm 0.005$ &
    $-0.435 \pm 0.006$ &
    $0.3460 \pm 0.0003$ &
    0.69 \\
NGC~4321 &
    $3.03 \pm 0.05$ &
    $-1.88 \pm 0.03$ &
    $0.412 \pm 0.003$ &
    $-0.385 \pm 0.004$ &
    $0.215 \pm 0.007$ &
    0.57 \\
NGC~4450 &
    $3.50 \pm 0.10$ &
    $-2.32 \pm 0.03$ &
    $0.370 \pm 0.005$ &
    $-0.431 \pm 0.006$ &
    $0.074 \pm 0.016$ &
    0.61 \\
NGC~4533 &
    $1.8 \pm 0.2$ &
    $-1.83 \pm 0.03$ &
    $0.796 \pm 0.012$ &
    $-0.099 \pm 0.007$ &
    $0.07 \pm 0.02$ &
    0.47 \\
NGC~4536 &
    $4.6 \pm 0.2$ &
    $-3.00 \pm 0.02$ &
    $0.249 \pm 0.003$ &
    $-0.604 \pm 0.006$ &
    $0.16 \pm 0.03$ &
    0.73 \\
NGC~4559 &
    $2.69 \pm 0.06$ &
    $-2.06 \pm 0.03$ &
    $0.333 \pm 0.002$ &
    $-0.477 \pm 0.003$ &
    $0.264 \pm 0.002$ &
    0.69 \\
NGC~4569 &
    $3.42 \pm 0.08$ &
    $-2.39 \pm 0.06$ &
    $0.342 \pm 0.003$ &
    $-0.466 \pm 0.003$ &
    $0.162 \pm 0.004$ &
    0.70 \\
NGC~4579 &
    $3.92 \pm 0.11$ &
    $-2.46 \pm 0.03$ &
    $0.338 \pm 0.004$ &
    $-0.471 \pm 0.005$ &
    $0.093 \pm 0.004$ &
    0.63 \\
NGC~4594 &
    $3.18 \pm 0.08$ &
    $-2.33 \pm 0.03$ &
    $0.348 \pm 0.003$ &
    $-0.458 \pm 0.004$ &
    $0.033 \pm 0.011$ &
    0.65 \\
NGC~4625 &
    $2.63 \pm 0.16$ &
    $-1.89 \pm 0.02$ &
    $0.343 \pm 0.011$ &
    $-0.465 \pm 0.014$ &
    $0.4658 \pm 0.0017$ &
    0.69 \\
NGC~4631 &
    $2.68 \pm 0.07$ &
    $-2.063 \pm 0.017$ &
    $0.2368 \pm 0.0016$ &
    $-0.626 \pm 0.003$ &
    $0.416 \pm 0.010$ &
    0.73 \\
NGC~4725 &
    $3.68 \pm 0.06$ &
    $-1.706 \pm 0.011$ &
    $0.342 \pm 0.002$ &
    $-0.466 \pm 0.003$ &
    $0.174 \pm 0.018$ &
    0.69 \\
NGC~4736 &
    $4.45 \pm 0.14$ &
    $-2.73 \pm 0.03$ &
    $0.130 \pm 0.002$ &
    $-0.886 \pm 0.007$ &
    $0.06 \pm 0.04$ &
    0.87 \\
NGC~4826 &
    $3.40 \pm 0.08$ &
    $-2.30 \pm 0.03$ &
    $0.284 \pm 0.003$ &
    $-0.546 \pm 0.004$ &
    $0.4374 \pm 0.0006$ &
    0.73 \\
NGC~5033 &
    $4.32 \pm 0.14$ &
    $-2.81 \pm 0.03$ &
    $0.181 \pm 0.002$ &
    $-0.743 \pm 0.006$ &
    $0.175 \pm 0.020$ &
    0.78 \\
NGC~5055 &
    $3.47 \pm 0.06$ &
    $-2.38 \pm 0.03$ &
    $0.307 \pm 0.002$ &
    $-0.512 \pm 0.003$ &
    $0.112 \pm 0.005$ &
    0.68 \\
\hline
\end{tabular}
\end{minipage}
\end{table*}

\begin{table*}
\centering
\begin{minipage}{147.5mm}
\renewcommand{\thefootnote}{\alph{footnote}}
\contcaption{}
\begin{tabular}{@{}lcccccc@{}}
\hline
    Name &
    $C$\footnotemark[1] &
    $\overline{M}_{20}$ &
    $\overline{R}_{eff}$\footnotemark[2] &
    $\log(\overline{R}_{eff})$\footnotemark[2] &
    $A$&
    $G$\\
\hline
NGC~5194 &
    $3.00 \pm 0.05$ &
    $-1.89 \pm 0.04$ &
    $0.439 \pm 0.002$ &
    $-0.358 \pm 0.002$ &
    $0.2517 \pm 0.0012$ &
    0.62 \\
NGC~5398 &
    $2.32 \pm 0.10$ &
    $-1.500 \pm 0.012$ &
    $0.293 \pm 0.009$ &
    $-0.534 \pm 0.013$ &
    $0.265 \pm 0.004$ &
    0.44 \\
NGC~5408 &
    $2.03 \pm 0.12$ &
    $-0.728 \pm 0.008$ &
    $0.555 \pm 0.015$ &
    $-0.256 \pm 0.012$ &
    $0.7188 \pm 0.0004$ &
    0.48 \\
NGC~5474 &
    $2.90 \pm 0.07$ &
    $-1.85 \pm 0.04$ &
    $0.499 \pm 0.005$ &
    $-0.302 \pm 0.005$ &
    $0.943 \pm 0.012$ &
    0.64 \\
NGC~5713 &
    $3.00 \pm 0.16$ &
    $-1.99 \pm 0.02$ &
    $0.292 \pm 0.009$ &
    $-0.534 \pm 0.013$ &
    $0.365 \pm 0.011$ &
    0.69 \\
NGC~5866 &
    $2.18 \pm 0.16$ &
    $-2.295 \pm 0.016$ &
    $0.481 \pm 0.005$ &
    $-0.318 \pm 0.005$ &
    $0.06 \pm 0.02$ &
    0.68 \\
NGC~6822 &
    $2.234 \pm 0.011$ &
    $-0.722 \pm 0.005$ &
    $0.6457 \pm 0.0016$ &
    $-0.1900 \pm 0.0011$ &
    $0.6917 \pm 0.0004$ &
    0.54 \\
NGC~6946 &
    $2.83 \pm 0.03$ &
    $-1.194 \pm 0.011$ &
    $0.519 \pm 0.002$ &
    $-0.2851 \pm 0.0018$ &
    $0.3956 \pm 0.0005$ &
    0.58 \\
NGC~7331 &
    $3.23 \pm 0.10$ &
    $-2.51 \pm 0.03$ &
    $0.326 \pm 0.002$ &
    $-0.486 \pm 0.003$ &
    $0.13068 \pm 0.00002$ &
    0.74 \\
NGC~7552 &
    $4.3 \pm 0.3$ &
    $-2.61 \pm 0.02$ &
    $0.217 \pm 0.007$ &
    $-0.664 \pm 0.015$ &
    $0.131 \pm 0.008$ &
    0.75 \\
NGC~7793 &
    $2.48 \pm 0.04$ &
    $-1.84 \pm 0.03$ &
    $0.496 \pm 0.003$ &
    $-0.305 \pm 0.002$ &
    $0.1890 \pm 0.0014$ &
    0.52 \\
\hline
\end{tabular}
$^a$ These data have been corrected for inclination effects using
     Equation~\ref{e_ccorr_inc}.\\
$^b$ These data have been corrected for inclination effects using
     Equation~\ref{e_lreffcorr_inc}.
\end{minipage}
\end{table*}

\begin{table*}
\centering
\begin{minipage}{145mm}
\renewcommand{\thefootnote}{\alph{footnote}}
\caption{Morphological Parameters Measured in 24~$\mu$m Data}
\label{t_param24}
\begin{tabular}{@{}lcccccc@{}}
\hline
    Name &
    $C$\footnotemark[1] &
    $\overline{M}_{20}$ &
    $\overline{R}_{eff}$\footnotemark[2] &
    $\log(\overline{R}_{eff})$\footnotemark[2] &
    $A$\footnotemark[3] &
    $G$ \\
\hline
DDO~53 &
    $4.9 \pm 0.4$ &
    $-2.67 \pm 0.03$ &
    $0.242 \pm 0.016$ &
    $-0.62 \pm 0.03$ &
    $0.70 \pm 0.03$ &
    0.73 \\
Ho~I &
    $1.32 \pm 0.04$ &
    $-1.11 \pm 0.10$ &
    $0.599 \pm 0.007$ &
    $-0.223 \pm 0.005$ &
    $3.28 \pm 0.11$ &
    0.45 \\
Ho~II &
    $2.42 \pm 0.04$ &
    $-0.946 \pm 0.012$ &
    $0.412 \pm 0.003$ &
    $-0.385 \pm 0.003$ &
    $1.456 \pm 0.006$ &
    0.68 \\
IC~2574 &
    $1.582 \pm 0.014$ &
    $-0.653 \pm 0.016$ &
    $0.4722 \pm 0.0019$ &
    $-0.3259 \pm 0.0017$ &
    $1.436 \pm 0.005$ &
    0.62 \\
IC~3583 &
    $2.13 \pm 0.15$ &
    $-1.075 \pm 0.013$ &
    $0.206 \pm 0.011$ &
    $-0.69 \pm 0.02$ &
    $1.586 \pm 0.005$ &
    0.67 \\
IC~4710 &
    $1.80 \pm 0.06$ &
    $-0.927 \pm 0.018$ &
    $0.392 \pm 0.007$ &
    $-0.406 \pm 0.008$ &
    $1.117 \pm 0.005$ &
    0.58 \\
Mrk~33 &
    $2.7 \pm 0.6$ &
    $-2.16 \pm 0.08$ &
    $0.20 \pm 0.03$ &
    $-0.70 \pm 0.05$ &
    $0.239 \pm 0.012$ &
    0.83 \\
NGC~24 &
    $2.02 \pm 0.10$ &
    $-1.30 \pm 0.03$ &
    $0.677 \pm 0.004$ &
    $-0.169 \pm 0.003$ &
    $0.506 \pm 0.003$ &
    0.64 \\
NGC~337 &
    $2.24 \pm 0.14$ &
    $-1.13 \pm 0.05$ &
    $0.429 \pm 0.009$ &
    $-0.368 \pm 0.009$ &
    $1.18 \pm 0.03$ &
    0.75 \\
NGC~586 &
    $1.8 \pm 0.2$ &
    $-1.09 \pm 0.02$ &
    $0.412 \pm 0.016$ &
    $-0.385 \pm 0.017$ &
    $0.289 \pm 0.009$ &
    0.65 \\
NGC~628 &
    $1.91 \pm 0.03$ &
    $-0.835 \pm 0.013$ &
    $0.386 \pm 0.002$ &
    $-0.414 \pm 0.003$ &
    $0.905 \pm 0.004$ &
    0.72 \\
NGC~855 &
    $2.6 \pm 0.4$ &
    $-2.10 \pm 0.03$ &
    $0.067 \pm 0.010$ &
    $-1.18 \pm 0.06$ &
    $0.30 \pm 0.03$ &
    0.83 \\
NGC~925 &
    $2.84 \pm 0.04$ &
    $-0.902 \pm 0.011$ &
    $0.520 \pm 0.002$ &
    $-0.284 \pm 0.002$ &
    $0.993 \pm 0.003$ &
    0.67 \\
NGC~1097 &
    $4.11 \pm 0.18$ &
    $-2.335 \pm 0.013$ &
    $0.068 \pm 0.003$ &
    $-1.166 \pm 0.017$ &
    $0.320 \pm 0.003$ &
    0.89 \\
NGC~1266 &
    $2.7 \pm 0.6$ &
    $-2.39 \pm 0.10$ &
    $0.182 \pm 0.016$ &
    $-0.74 \pm 0.04$ &
    $0.34 \pm 0.02$ &
    0.89 \\
NGC~1291 &
    $5.55 \pm 0.09$ &
    $-1.60 \pm 0.05$ &
    $0.481 \pm 0.003$ &
    $-0.318 \pm 0.002$ &
    $0.842 \pm 0.003$ &
    0.61 \\
NGC~1316 &
    $5.65 \pm 0.18$ &
    $-3.63 \pm 0.03$ &
    $0.141 \pm 0.002$ &
    $-0.852 \pm 0.006$ &
    $1.06 \pm 0.04$ &
    0.70 \\
NGC~1317 &
    $2.43 \pm 0.19$ &
    $-1.700 \pm 0.015$ &
    $0.201 \pm 0.009$ &
    $-0.70 \pm 0.02$ &
    $0.254 \pm 0.006$ &
    0.80 \\
NGC~1377 &
    $2.4 \pm 0.6$ &
    $-2.26 \pm 0.08$ &
    $0.160 \pm 0.014$ &
    $-0.79 \pm 0.04$ &
    $0.25 \pm 0.05$ &
    0.89 \\
NGC~1482 &
    $2.4 \pm 0.5$ &
    $-2.26 \pm 0.05$ &
    $0.147 \pm 0.010$ &
    $-0.83 \pm 0.03$ &
    $0.196 \pm 0.014$ &
    0.90 \\
NGC~1510 &
    $2.6 \pm 0.6$ &
    $-2.06 \pm 0.06$ &
    $0.279 \pm 0.019$ &
    $-0.55 \pm 0.03$ &
    $0.225 \pm 0.013$ &
    0.83 \\
NGC~1512 &
    $4.83 \pm 0.18$ &
    $-2.98 \pm 0.02$ &
    $0.318 \pm 0.003$ &
    $-0.497 \pm 0.004$ &
    $0.62 \pm 0.02$ &
    0.76 \\
NGC~1566 &
    $2.25 \pm 0.06$ &
    $-1.045 \pm 0.009$ &
    $0.273 \pm 0.003$ &
    $-0.564 \pm 0.005$ &
    $0.777 \pm 0.018$ &
    0.85 \\
NGC~1705 &
    $2.06 \pm 0.19$ &
    $-1.289 \pm 0.011$ &
    $0.340 \pm 0.013$ &
    $-0.468 \pm 0.017$ &
    $0.319 \pm 0.013$ &
    0.72 \\
NGC~2403 &
    $2.31 \pm 0.03$ &
    $-1.138 \pm 0.015$ &
    $0.2443 \pm 0.0011$ &
    $-0.612 \pm 0.002$ &
    $0.769 \pm 0.002$ &
    0.82 \\
NGC~2798 &
    $2.5 \pm 0.6$ &
    $-2.31 \pm 0.05$ &
    $0.138 \pm 0.010$ &
    $-0.86 \pm 0.03$ &
    $0.20 \pm 0.04$ &
    0.90 \\
NGC~2841 &
    $1.65 \pm 0.04$ &
    $-1.00 \pm 0.02$ &
    $0.448 \pm 0.003$ &
    $-0.348 \pm 0.003$ &
    $0.2712 \pm 0.0010$ &
    0.61 \\
NGC~2915 &
    $2.8 \pm 0.3$ &
    $-1.503 \pm 0.016$ &
    $0.171 \pm 0.013$ &
    $-0.77 \pm 0.03$ &
    $0.587 \pm 0.003$ &
    0.67 \\
NGC~2976 &
    $1.29 \pm 0.05$ &
    $-0.636 \pm 0.009$ &
    $0.764 \pm 0.004$ &
    $-0.117 \pm 0.002$ &
    $0.5903 \pm 0.0011$ &
    0.77 \\
NGC~3031 &
    $2.68 \pm 0.02$ &
    $-1.01 \pm 0.02$ &
    $0.4560 \pm 0.0009$ &
    $-0.3411 \pm 0.0009$ &
    $0.581 \pm 0.005$ &
    0.71 \\
NGC~3049 &
    $3.3 \pm 0.6$ &
    $-2.48 \pm 0.06$ &
    $0.139 \pm 0.011$ &
    $-0.86 \pm 0.04$ &
    $0.27 \pm 0.07$ &
    0.85 \\
NGC~3184 &
    $2.04 \pm 0.04$ &
    $-0.987 \pm 0.018$ &
    $0.428 \pm 0.003$ &
    $-0.368 \pm 0.003$ &
    $0.691 \pm 0.006$ &
    0.66 \\
NGC~3185 &
    $4.5 \pm 0.4$ &
    $-2.61 \pm 0.04$ &
    $0.199 \pm 0.011$ &
    $-0.70 \pm 0.02$ &
    $0.270 \pm 0.015$ &
    0.76 \\
NGC~3187 &
    $2.27 \pm 0.14$ &
    $-1.408 \pm 0.016$ &
    $0.706 \pm 0.008$ &
    $-0.151 \pm 0.005$ &
    $0.865 \pm 0.004$ &
    0.75 \\
NGC~3190 &
    $2.8 \pm 0.2$ &
    $-2.51 \pm 0.04$ &
    $0.488 \pm 0.006$ &
    $-0.312 \pm 0.005$ &
    $0.4161 \pm 0.0012$ &
    0.79 \\
NGC~3198 &
    $5.0 \pm 0.2$ &
    $-1.41 \pm 0.09$ &
    $0.286 \pm 0.003$ &
    $-0.543 \pm 0.004$ &
    $0.48 \pm 0.05$ &
    0.80 \\
NGC~3265 &
    $2.9 \pm 0.6$ &
    $-2.23 \pm 0.05$ &
    $0.136 \pm 0.019$ &
    $-0.87 \pm 0.06$ &
    $0.15 \pm 0.15$ &
    0.85 \\
NGC~3351 &
    $5.2 \pm 0.3$ &
    $-2.663 \pm 0.008$ &
    $0.074 \pm 0.003$ &
    $-1.13 \pm 0.02$ &
    $0.261 \pm 0.018$ &
    0.83 \\
NGC~3521 &
    $2.15 \pm 0.06$ &
    $-1.248 \pm 0.008$ &
    $0.248 \pm 0.002$ &
    $-0.605 \pm 0.004$ &
    $0.3735 \pm 0.0011$ &
    0.82 \\
NGC~3621 &
    $1.97 \pm 0.05$ &
    $-1.143 \pm 0.009$ &
    $0.278 \pm 0.002$ &
    $-0.556 \pm 0.003$ &
    $0.644 \pm 0.009$ &
    0.84 \\
NGC~3627 &
    $1.12 \pm 0.04$ &
    $-0.920 \pm 0.013$ &
    $0.378 \pm 0.003$ &
    $-0.422 \pm 0.003$ &
    $0.851 \pm 0.013$ &
    0.83 \\
NGC~3773 &
    $2.5 \pm 0.5$ &
    $-2.06 \pm 0.05$ &
    $0.22 \pm 0.02$ &
    $-0.66 \pm 0.04$ &
    $0.21 \pm 0.03$ &
    0.82 \\
\hline
\end{tabular}
\end{minipage}
\end{table*}

\begin{table*}
\centering
\begin{minipage}{145mm}
\renewcommand{\thefootnote}{\alph{footnote}}
\contcaption{}
\begin{tabular}{@{}lcccccc@{}}
\hline
    Name &
    $C$\footnotemark[1] &
    $\overline{M}_{20}$ &
    $\overline{R}_{eff}$\footnotemark[2] &
    $\log(\overline{R}_{eff})$\footnotemark[2] &
    $A$\footnotemark[3]&
    $G$\\
\hline
NGC~3938 &
    $2.36 \pm 0.06$ &
    $-0.857 \pm 0.009$ &
    $0.392 \pm 0.005$ &
    $-0.407 \pm 0.005$ &
    $0.692 \pm 0.003$ &
    0.71 \\
NGC~4125 &
    $4.40 \pm 0.16$ &
    $-2.76 \pm 0.03$ &
    $0.148 \pm 0.004$ &
    $-0.831 \pm 0.013$ &
    $0.868 \pm 0.011$ &
    0.64 \\
NGC~4236 &
    $1.414 \pm 0.013$ &
    $-0.896 \pm 0.014$ &
    $0.2514 \pm 0.0011$ &
    $-0.600 \pm 0.002$ &
    $1.361 \pm 0.003$ &
    0.67 \\
NGC~4254 &
    $2.69 \pm 0.07$ &
    $-1.50 \pm 0.02$ &
    $0.379 \pm 0.005$ &
    $-0.421 \pm 0.005$ &
    $0.756 \pm 0.003$ &
    0.74 \\
NGC~4321 &
    $4.54 \pm 0.12$ &
    $-2.23 \pm 0.11$ &
    $0.344 \pm 0.003$ &
    $-0.464 \pm 0.004$ &
    $0.675 \pm 0.005$ &
    0.75 \\
NGC~4450 &
    $3.62 \pm 0.12$ &
    $-1.98 \pm 0.04$ &
    $0.363 \pm 0.005$ &
    $-0.440 \pm 0.006$ &
    $0.77 \pm 0.03$ &
    0.65 \\
NGC~4533 &
    $1.2 \pm 0.2$ &
    $-1.364 \pm 0.012$ &
    $0.690 \pm 0.012$ &
    $-0.161 \pm 0.008$ &
    $0.298 \pm 0.013$ &
    0.56 \\
NGC~4536 &
    $3.8 \pm 0.4$ &
    $-2.88 \pm 0.03$ &
    $0.077 \pm 0.003$ &
    $-1.112 \pm 0.019$ &
    $0.29 \pm 0.08$ &
    0.90 \\
NGC~4559 &
    $2.48 \pm 0.06$ &
    $-1.075 \pm 0.010$ &
    $0.295 \pm 0.002$ &
    $-0.529 \pm 0.003$ &
    $0.8205 \pm 0.0009$ &
    0.77 \\
NGC~4569 &
    $5.3 \pm 0.4$ &
    $-3.278 \pm 0.018$ &
    $0.142 \pm 0.003$ &
    $-0.848 \pm 0.008$ &
    $0.56 \pm 0.04$ &
    0.89 \\
NGC~4579 &
    $4.10 \pm 0.15$ &
    $-1.803 \pm 0.018$ &
    $0.362 \pm 0.004$ &
    $-0.441 \pm 0.005$ &
    $0.57 \pm 0.02$ &
    0.74 \\
NGC~4594 &
    $3.54 \pm 0.08$ &
    $-1.17 \pm 0.05$ &
    $0.433 \pm 0.003$ &
    $-0.363 \pm 0.003$ &
    $0.284 \pm 0.019$ &
    0.72 \\
NGC~4618 &
    $2.26 \pm 0.06$ &
    $-1.075 \pm 0.018$ &
    $0.342 \pm 0.006$ &
    $-0.466 \pm 0.008$ &
    $0.8625 \pm 0.0011$ &
    0.64 \\
NGC~4625 &
    $2.44 \pm 0.16$ &
    $-1.564 \pm 0.015$ &
    $0.297 \pm 0.011$ &
    $-0.527 \pm 0.017$ &
    $0.688 \pm 0.007$ &
    0.74 \\
NGC~4631 &
    $2.77 \pm 0.07$ &
    $-1.817 \pm 0.016$ &
    $0.1757 \pm 0.0016$ &
    $-0.755 \pm 0.004$ &
    $0.67 \pm 0.02$ &
    0.81 \\
NGC~4725 &
    $1.39 \pm 0.02$ &
    $-0.98 \pm 0.03$ &
    $0.455 \pm 0.002$ &
    $-0.342 \pm 0.002$ &
    $0.9781 \pm 0.0016$ &
    0.67 \\
NGC~4736 &
    $2.93 \pm 0.10$ &
    $-1.661 \pm 0.017$ &
    $0.145 \pm 0.002$ &
    $-0.839 \pm 0.007$ &
    $0.25 \pm 0.02$ &
    0.89 \\
NGC~4826 &
    $2.80 \pm 0.16$ &
    $-2.172 \pm 0.015$ &
    $0.113 \pm 0.003$ &
    $-0.948 \pm 0.010$ &
    $0.9207 \pm 0.0009$ &
    0.91 \\
NGC~5033 &
    $3.95 \pm 0.12$ &
    $-2.45 \pm 0.03$ &
    $0.175 \pm 0.002$ &
    $-0.757 \pm 0.006$ &
    $0.5065 \pm 0.0005$ &
    0.82 \\
NGC~5055 &
    $2.78 \pm 0.05$ &
    $-1.240 \pm 0.015$ &
    $0.319 \pm 0.002$ &
    $-0.496 \pm 0.003$ &
    $0.4326 \pm 0.0008$ &
    0.78 \\
NGC~5194 &
    $3.02 \pm 0.05$ &
    $-1.246 \pm 0.011$ &
    $0.443 \pm 0.002$ &
    $-0.354 \pm 0.002$ &
    $0.5603 \pm 0.0009$ &
    0.73 \\
NGC~5398 &
    $0.77 \pm 0.08$ &
    $-0.78 \pm 0.03$ &
    $0.328 \pm 0.009$ &
    $-0.484 \pm 0.012$ &
    $1.9184 \pm 0.0019$ &
    0.85 \\
NGC~5408 &
    $0.69 \pm 0.08$ &
    $-0.64 \pm 0.03$ &
    $0.647 \pm 0.015$ &
    $-0.189 \pm 0.010$ &
    $1.4430 \pm 0.0008$ &
    0.85 \\
NGC~5474 &
    $2.66 \pm 0.05$ &
    $-1.09 \pm 0.03$ &
    $0.592 \pm 0.005$ &
    $-0.228 \pm 0.004$ &
    $1.056 \pm 0.004$ &
    0.58 \\
NGC~5713 &
    $3.1 \pm 0.3$ &
    $-1.719 \pm 0.010$ &
    $0.184 \pm 0.009$ &
    $-0.73 \pm 0.02$ &
    $1.240 \pm 0.013$ &
    0.87 \\
NGC~5866 &
    $1.8 \pm 0.2$ &
    $-1.89 \pm 0.02$ &
    $0.329 \pm 0.005$ &
    $-0.483 \pm 0.007$ &
    $0.21 \pm 0.04$ &
    0.83 \\
NGC~6822 &
    $0.969 \pm 0.008$ &
    $-0.951 \pm 0.010$ &
    $0.5762 \pm 0.0016$ &
    $-0.2394 \pm 0.0012$ &
    $0.9146 \pm 0.0018$ &
    0.73 \\
NGC~6946 &
    $5.80 \pm 0.13$ &
    $-2.02 \pm 0.11$ &
    $0.380 \pm 0.002$ &
    $-0.420 \pm 0.002$ &
    $0.562 \pm 0.014$ &
    0.78 \\
NGC~7331 &
    $2.22 \pm 0.07$ &
    $-1.469 \pm 0.010$ &
    $0.397 \pm 0.002$ &
    $-0.401 \pm 0.003$ &
    $0.448 \pm 0.005$ &
    0.80 \\
NGC~7552 &
    $3.1 \pm 0.4$ &
    $-2.47 \pm 0.04$ &
    $0.093 \pm 0.007$ &
    $-1.03 \pm 0.03$ &
    $0.27 \pm 0.02$ &
    0.91 \\
NGC~7793 &
    $2.20 \pm 0.04$ &
    $-1.054 \pm 0.017$ &
    $0.469 \pm 0.003$ &
    $-0.329 \pm 0.002$ &
    $0.5105 \pm 0.0005$ &
    0.65 \\
\hline
\end{tabular}
$^a$ These data have been corrected for inclination effects using
     Equation~\ref{e_ccorr_inc}.\\
$^b$ These data have been corrected for inclination effects using
     Equation~\ref{e_lreffcorr_inc}.\\
$^c$ These data have been corrected for distance-related 
     effects using Equation~\ref{e_acorr_dist}.\\
\end{minipage}
\end{table*}
\renewcommand{\thefootnote}{\arabic{footnote}}

This analysis demonstrates that the $C$ parameter has the added
disadvantage over the $\overline{M}_{20}$ parameter of being more
sensitive to inclination effects.  Given that $\overline{M}_{20}$ is
also a more sensitive concentration parameter \citep{lpm04}, it
should be used instead of $C$ as a concentration parameter.  The $C$
parameter is still included here so as to allow for better comparisons
between this paper's results and the results of \citet{c03}.

\begin{table*}
\centering
\begin{minipage}{130mm}
\renewcommand{\thefootnote}{\alph{footnote}}
\caption{Statistics on 3.6~$\mu$m Morphological Parameters for 
    Different Hubble Types}
\label{t_paramirac1_htype}
\begin{tabular}{@{}lcccccc@{}}
\hline
    Hubble &
    Number of &
    $C$\footnotemark[1] &
    $\overline{M}_{20}$\footnotemark[1] &
    $\log(\overline{R}_{eff})$\footnotemark[1] &
    $A$\footnotemark[1] &
    $G$\footnotemark[1] \\
    Type &
    Galaxies &
    &
    &
    &
    &
    \\
\hline
All &     71 &
    $3.00 \pm 0.09$ &
    $-2.03 \pm 0.08$ &
    $-0.47 \pm 0.02$ &
    $0.18 \pm 0.03$ &    
    $0.637 \pm 0.017$ \\
E-S0 &    11 &    
    $3.1 \pm 0.2$ &
    $-2.20 \pm 0.07$ &
    $-0.49 \pm 0.06$ &
    $0.09 \pm 0.03$ &    
    $0.65 \pm 0.02$ \\
S0/a-Sab &  15 &
    $3.50 \pm 0.17$ &
    $-2.37 \pm 0.07$ &
    $-0.50 \pm 0.04$ &
    $0.12 \pm 0.02$ &    
    $0.70 \pm 0.02$ \\
Sb-Sbc &  13 &
    $3.37 \pm 0.17$ &
    $-2.41 \pm 0.09$ &
    $-0.49 \pm 0.02$ &
    $0.16 \pm 0.03$ &    
    $0.682 \pm 0.015$ \\
Sc-Sd &   19 &
    $2.71 \pm 0.13$ &
    $-1.73 \pm 0.09$ &
    $-0.36 \pm 0.04$ &
    $0.27 \pm 0.04$ &    
    $0.630 \pm 0.019$ \\
Sdm-Im &  13 &
    $2.44 \pm 0.11$ &
    $-0.88 \pm 0.14$ &
    $-0.46 \pm 0.05$ &
    $0.69 \pm 0.08$ &    
    $0.48 \pm 0.03$ \\
\hline
\end{tabular}
$^a$ These are median values and standard deviations of the means.
\end{minipage}
\end{table*}

\begin{table*}
\centering
\begin{minipage}{130mm}
\renewcommand{\thefootnote}{\alph{footnote}}
\caption{Statistics on 24~$\mu$m Morphological Parameters for 
    Different Hubble Types}
\label{t_param24_htype}
\begin{tabular}{@{}lcccccc@{}}
\hline
    Hubble &
    Number of &
    $C$\footnotemark[1] &
    $\overline{M}_{20}$\footnotemark[1] &
    $\log(\overline{R}_{eff})$\footnotemark[1] &
    $A$\footnotemark[1] &
    $G$\footnotemark[1] \\
    Type &
    Galaxies &
    &
    &
    &
    &
    \\
\hline
All &     73 &
    $2.51 \pm 0.14$ &
    $-1.47 \pm 0.08$ &
    $-0.50 \pm 0.03$ &    
    $0.59 \pm 0.06$ &
    $0.764 \pm 0.012$ \\
E-S0 &    11 &    
    $2.6 \pm 0.3$ &
    $-2.23 \pm 0.17$ &
    $-0.80 \pm 0.06$ &
    $0.25 \pm 0.09$ &    
    $0.83 \pm 0.03$ \\
S0/a-Sab &  16 &
    $3.0 \pm 0.3$ &
    $-2.07 \pm 0.18$ &
    $-0.60 \pm 0.06$ &
    $0.35 \pm 0.07$ &    
    $0.77 \pm 0.03$ \\
Sb-Sbc &  13 &
    $3.0 \pm 0.3$ &
    $-1.47 \pm 0.18$ &
    $-0.50 \pm 0.08$ &
    $0.45 \pm 0.08$ &    
    $0.82 \pm 0.02$ \\
Sc-Sd &   19 &
    $2.3 \pm 0.3$ &
    $-1.14 \pm 0.10$ &
    $-0.41 \pm 0.04$ &
    $0.69 \pm 0.05$ &    
    $0.745 \pm 0.019$ \\
Sdm-Im &  14 &
    $2.0 \pm 0.3$ &
    $-1.01 \pm 0.15$ &
    $-0.48 \pm 0.05$ &
    $1.24 \pm 0.20$ &    
    $0.68 \pm 0.03$ \\
\hline
\end{tabular}
$^a$ These are median values and standard deviations of the means.
\end{minipage}
\end{table*}

\begin{figure*}
\epsfig{file=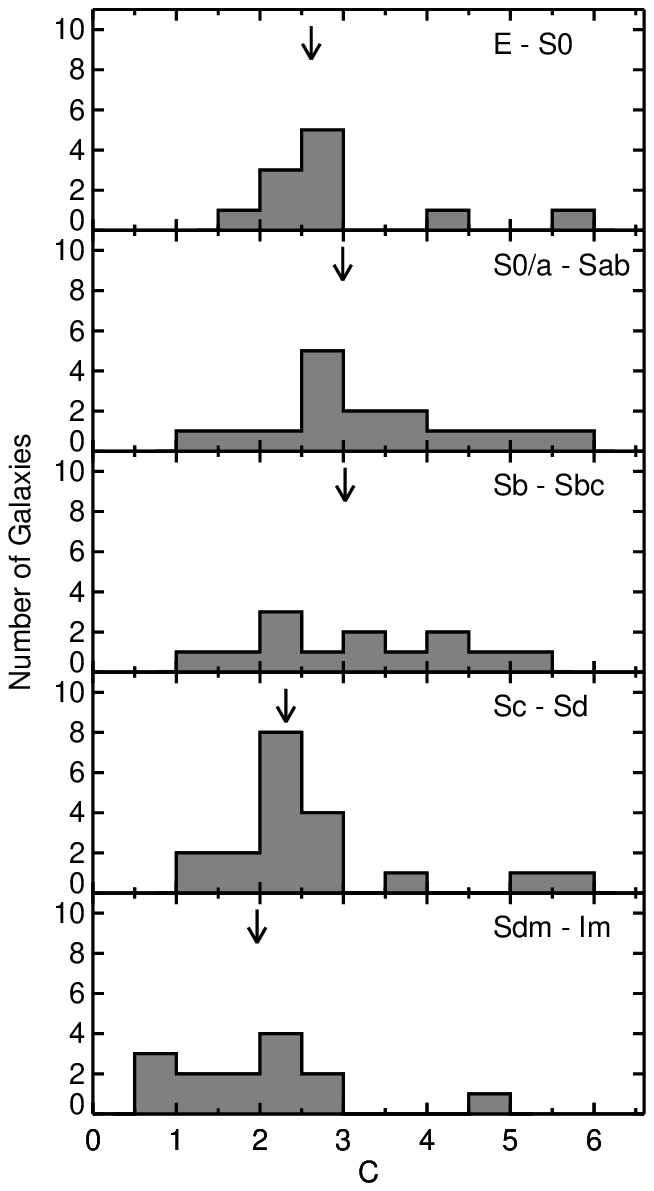, height=115mm}\epsfig{file=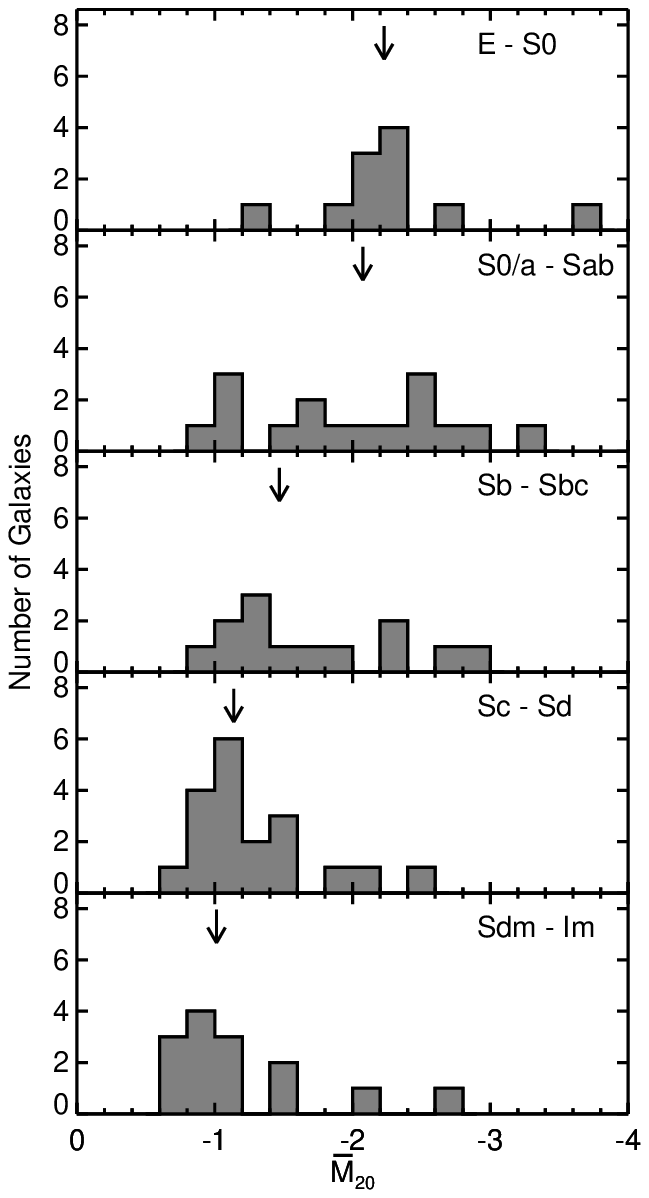, height=115mm}
\caption{Histograms of the 24~$\mu$m $C$ and $\overline{M}_{20}$
parameters sorted according to Hubble type.  The parameters are
plotted so that more concentrated (higher $C$, lower
$\overline{M}_{20}$) values appear on the right.  The arrow indicates
the median value for each subgroup (with the average of the two central
values used to calculate the median in subgroups with an even number
of members).  See Section~\ref{s_morphparam_tanal}
for discussion.}
\label{f_cm20_vs_t}
\end{figure*}

\section{Trends in the Morphological Parameters with Hubble Type}
    \label{s_morphparam_ttrend}

\subsection{Statistical Analysis of the Trends in the Morphological Parameters}
    \label{s_morphparam_tanal}

Table~\ref{t_paramirac1} gives the morphological parameters measured
from the 3.6~$\mu$m data, and Table~\ref{t_param24} gives the
parameters measured from the 24~$\mu$m data.  Medians and standard
deviations of the means for the parameters as a function of Hubble
type are given in Tables~\ref{t_paramirac1_htype} and
\ref{t_param24_htype}.  To examine the variations versus morphology,
similar Hubble types have been grouped together (i.e. S0/a-Sab;
Sb-Sbc, etc.).  As an additional test of the trends in the
morphological parameters, we have applied the Kolmogorov-Smirnov (K-S)
test, which gives the probability that individual parameters for the
S0/a-Sab and Sc-Sd galaxies are drawn from the same distribution.  If
this probability is lower than 1\%, the S0/a-Sab values are
statistically different from the Sc-Sd values.  (A 1\% probability
that two populations are drawn from the same distribution should be
equivalent to a $3\sigma$ difference between the median values of the
two populations.)  The results of this test on the 24~$\mu$m
morphological parameters is given in Table~\ref{t_kstest_htype}.

Although both the 3.6 and 24~$\mu$m morphological parameters are
presented here, the emphasis of the analysis in this section is placed
on the 24~$\mu$m morphological parameters, since variations in stellar
morphologies along the Hubble sequence are already well-understood.
We note that the 3.6~$\mu$m morphological parameters show that
starlight is more centrally concentrated and more symmetric in
early-type galaxies and more extended and asymmetric in late-type
galaxies, as is expected from the definition of the Hubble sequence
\citep{h26}.  Moreover, the parameters derived here match the R-band
morphological parameters derived for normal spiral galaxies by
\citet{c03} and \citet{lpm04}.  However, we save
the detailed discussion about the 3.6~$\mu$m parameters for
Section~\ref{s_discuss_compstellar}, where the stellar morphological
parameters are compared to the 24~$\mu$m morphological parameters.
The rest of this section is used to search for possible trends in the
24~$\mu$m morphological parameters along the Hubble sequence.

\begin{table}
\begin{center}
\caption{Results on Applying K-S Test to 24~$\mu$m Morphological
    Parameters for S0/a-Sab and Sc-Sd Galaxies}
\label{t_kstest_htype}
\begin{tabular}{@{}cc@{}}
\hline
    Parameter &    Probability\footnotemark[1] \\
\hline
    $C$ &                              0.025 \\
    $\overline{M}_{20}$ &       0.0023 \\
    $\log(\overline{R}_{eff})$ & 0.099 \\
    $A$ &                              0.013 \\
    $G$ &                              0.27 \\
\hline
\end{tabular}
\end{center}
$^a$ This is the probability that the tested parameter for the
     S0/a-Sab and Sc-Sd data have the same distribution.  A
     probability of $\sim0.01$ is interpreted as having the same
     significance as a $3\sigma$ difference between the median values
     for the S0/a-Sab and Sc-Sd galaxies.
\end{table}
\renewcommand{\thefootnote}{\arabic{footnote}}

\begin{figure}
\begin{center}
\epsfig{file=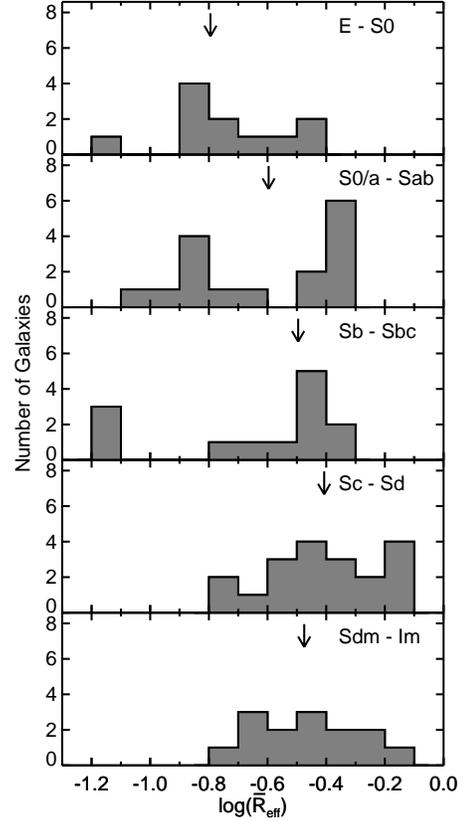, height=115mm}
\end{center}
\caption{Histograms of the 24~$\mu$m $\log(\overline{R}_{eff})$
parameter sorted according to Hubble type.  Higher
$\log(\overline{R}_{eff})$ values correspond to dust emission that is
more extended relative to the optical disc.  The arrow indicates the
median value for each subgroup (with the average of the two central
values used to calculate the median in subgroups with an even number
of members, which is why the median does not appear to correspond to
any galaxy in one of the panels).  See
Section~\ref{s_morphparam_tanal} for discussion.}
\label{f_reff_vs_t}
\end{figure}

\begin{figure}
\begin{center}
\epsfig{file=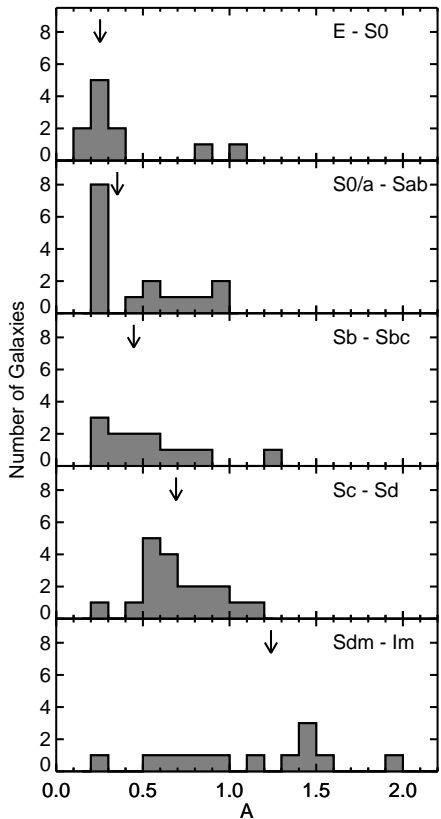, height=115mm}
\end{center}
\caption{Histograms of the 24~$\mu$m $A$ parameter sorted according to
Hubble type.  Higher $A$ values correspond to higher asymmetry.  Note
that the Im galaxy Ho~I, with $A=3.28\pm0.11$, falls to the right of
the plot range in the bottom histogram.  The arrow indicates the
median value for each subgroup (with the average of the two central
values used to calculate the median in subgroups with an even number
of members, which is why the median does not appear to correspond to
any galaxy in two of the panels).  See
Section~\ref{s_morphparam_tanal} for discussion.}
\label{f_a_vs_t}
\end{figure}

\begin{figure}
\begin{center}
\epsfig{file=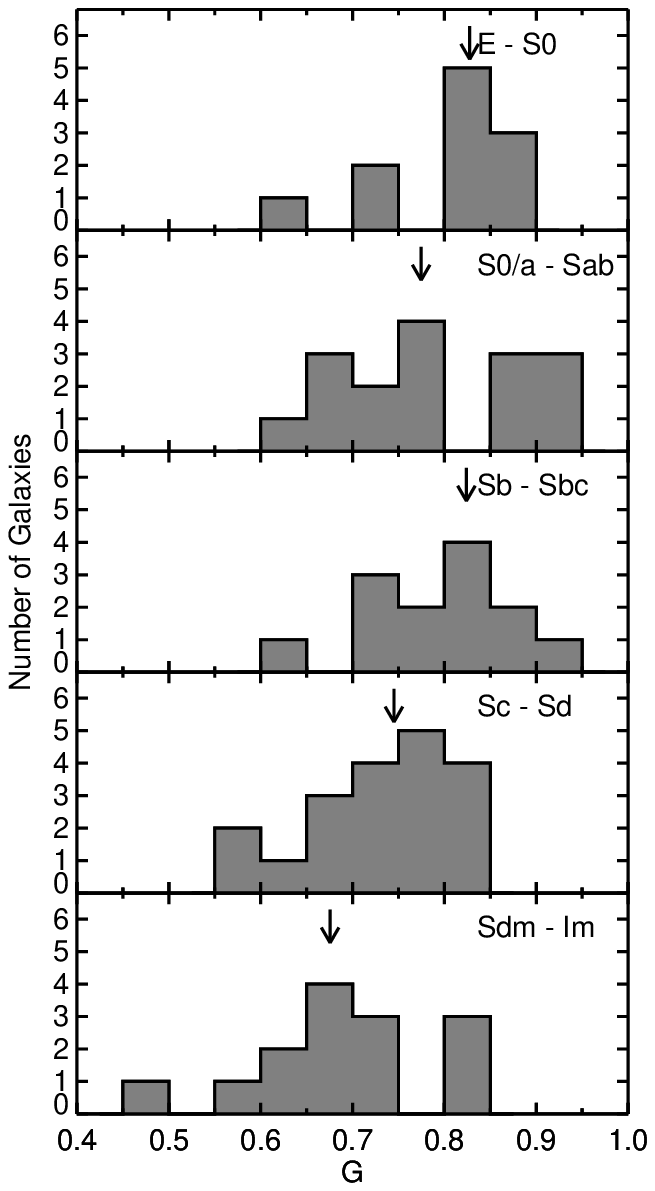, height=115mm}
\end{center}
\caption{Histograms of the 24~$\mu$m $G$ parameter sorted according to
Hubble type.  Higher $G$ values correspond to flux being concentrated
in a few bright knots; lower $G$ values correspond to dust emission
being more evenly distributed.  The arrow indicates the median value
for each subgroup (with the average of the two central
values used to calculate the median in subgroups with an even number
of members). See Section~\ref{s_morphparam_tanal} for discussion.}
\label{f_g_vs_t}
\end{figure}

Figure~\ref{f_cm20_vs_t} shows how the 24~$\mu$m $C$ and
$\overline{M}_{20}$ parameters vary with morphological type.
The $C$ parameter only appears to vary slightly between S0/a-Sab and
Sc-Sd galaxies, but it is significantly lower in Sdm-Im galaxies.  The
weak variations in the $C$ parameter may partly reflect its
ineffectiveness as a measure of central concentration.  Stronger
trends are visible in the $\overline{M}_{20}$ parameter as a
function of Hubble type, where a strong difference can be seen between
S0/a-Sab and Sc-Sd galaxies.  These trends are also reflected in both
the statistics in Table~\ref{t_param24_htype} and the results from the
K-S test in Table~\ref{t_kstest_htype}.  The results demonstrate that
the central concentration of 24~$\mu$m emission varies across the
Hubble sequence, with the emission being the most centrally
concentrated in E-S0 galaxies and the most diffuse in Sdm-Im galaxies.
Note the statistically significant variations between S0/a-Sab and
Sc-Sd galaxies.

Figure~\ref{f_reff_vs_t} shows how the 24~$\mu$m
$\log(\overline{R}_{eff})$ parameter varies with morphological
type.  Along with the statistical information in
Tables~\ref{t_param24_htype} and \ref{t_kstest_htype},
Figure~\ref{f_reff_vs_t} shows general trends in
$\log(\overline{R}_{eff})$ as a function of Hubble type, with
the most centralized infrared emission found in E-S0 galaxies and the
most extended emission found in Sdm-Im galaxies.  However, the
statistical significance of the difference between the distribution of
$\log(\overline{R}_{eff})$ for S0/a-Sab and Sc-Sd galaxies is
relatively weak.  The difference between the values in
Table~\ref{t_param24_htype} is less than $3\sigma$, and the K-S test
indicates that the probability that the S0/a-Sab and Sc-Sd data come
from the same distribution is $\sim10$\% (whereas a 1\% probability
would be the equivalent of a $3\sigma$ difference).  Despite the weak
statistical significance of the difference between S0/a-Sab and Sc-Sd
galaxies, these results still indicate that the 24~$\mu$m emission is
relatively centralized in early-type spiral galaxies but relatively
extended in late-type spiral galaxies.

Figure~\ref{f_a_vs_t} shows how the 24~$\mu$m $A$ parameter varies with
morphological type.  A trend with morphological type is seen with this
parameter as well, with E-S0 galaxies showing the most symmetric
emission and Sdm-Im galaxies showing the most asymmetric emission.
Again, the results in Tables~\ref{t_param24_htype} and
\ref{t_kstest_htype} show that statistically significant differences
can be seen between the S0/a-Sab and Sc-Sd galaxies, although the
variations are close to the $3\sigma$ level.  Also note the
relatively large dispersion in values for Sdm-Im galaxies.

Figure~\ref{f_g_vs_t} shows how the 24~$\mu$m $G$ parameter varies
with morphological type.  The variations in this parameter are weak.
Statistically, no significant change is observed between E and Sd
galaxies, although some Sc-Sd galaxies have $G$ values that are lower
than the G parameters for most E-Sbc galaxies.  This may indicate that
the same fraction of 24~$\mu$m emission originates from bright peaks
in most early- and late-type spiral galaxies, regardless of how the
bright peaks are distributed within the galaxies, and that only a few
late-type spiral galaxies contain relatively more high surface
brightness 24~$\mu$m emission from diffuse or extended sources.
Sdm-Im galaxies typically have slightly lower G values than other
galaxies, indicating that the infrared emission is more evenly
distributed than other galaxy types.  However strong variations in $G$
among the Sdm-Im galaxies are visible, which indicates that some
Sdm-Im galaxies have very evenly-distributed emission while others
have point-like emission.

Overall, these results reveal a clear trend in the distribution of
24~$\mu$m dust emission along the Hubble sequence.  24~$\mu$m emission
is generally compact (relative to itself and relative to stellar
emission) and symmetric in E-S0 galaxies.  In S0/a-Sab galaxies, the
dust emission is only slightly more extended; the discs are generally
only a weak source of emission.  In Sc-Sd galaxies, nuclear dust
emission is still present, but a much greater fraction of the emission
originates from the discs of the galaxies, as reflected by the high
$\overline{M}_{20}$ and $\log(\overline{R}_{eff})$
values.  The dust emission in the discs also appears to be more
asymmetric.  Finally, in Sdm-Im galaxies, the galaxies' structure is
disorganized, resulting in highly extended, highly asymmetric dust
emission.

Some of the Sdm-Im galaxies are exceptions to the trends seen with the
other galaxies in this sample.  The 24~$\mu$m
$\overline{M}_{20}$, $\log(\overline{R}_{eff})$, $A$, and $G$
parameters for these galaxies span a broad range of values.  Some
objects, such as DDO~53 and Mrk~33, are compact sources, while other
galaxies, such as NGC~5408 and Ho~II, are highly extended, asymmetric
objects.  Also worth noting are the Im galaxies not included in this
analysis because they were effectively non-detections in one or more
MIPS bands, including DDO~154, DDO~165, Ho~IX, M81 Dwarf A, and M81
Dwarf B.  This is discussed further in Section~\ref{s_discuss_irr}.

\subsection{Galaxies with Representative 24~$\mu$m Morphologies}
    \label{s_morphparam_rep}

Based on the 24~$\mu$m parameters, we can compare galaxies of a given
Hubble type and determine which galaxies are representative of each
type in the 24~$\mu$m wave band.  We show the 24~$\mu$m images of
these representative galaxies in Figure~\ref{f_represent}.  The
galaxies in Figure~\ref{f_represent} were selected on the basis that
their morphological parameters are statistically equivalent to the
median morphological parameters for the given Hubble type.  They
sometimes contrast with nearby or well-studied galaxies where the
structure of the 24~$\mu$m dust emission (and possibly the structure
of the ISM) can be very peculiar for its particular Hubble type.

\begin{figure*}
\epsfig{file=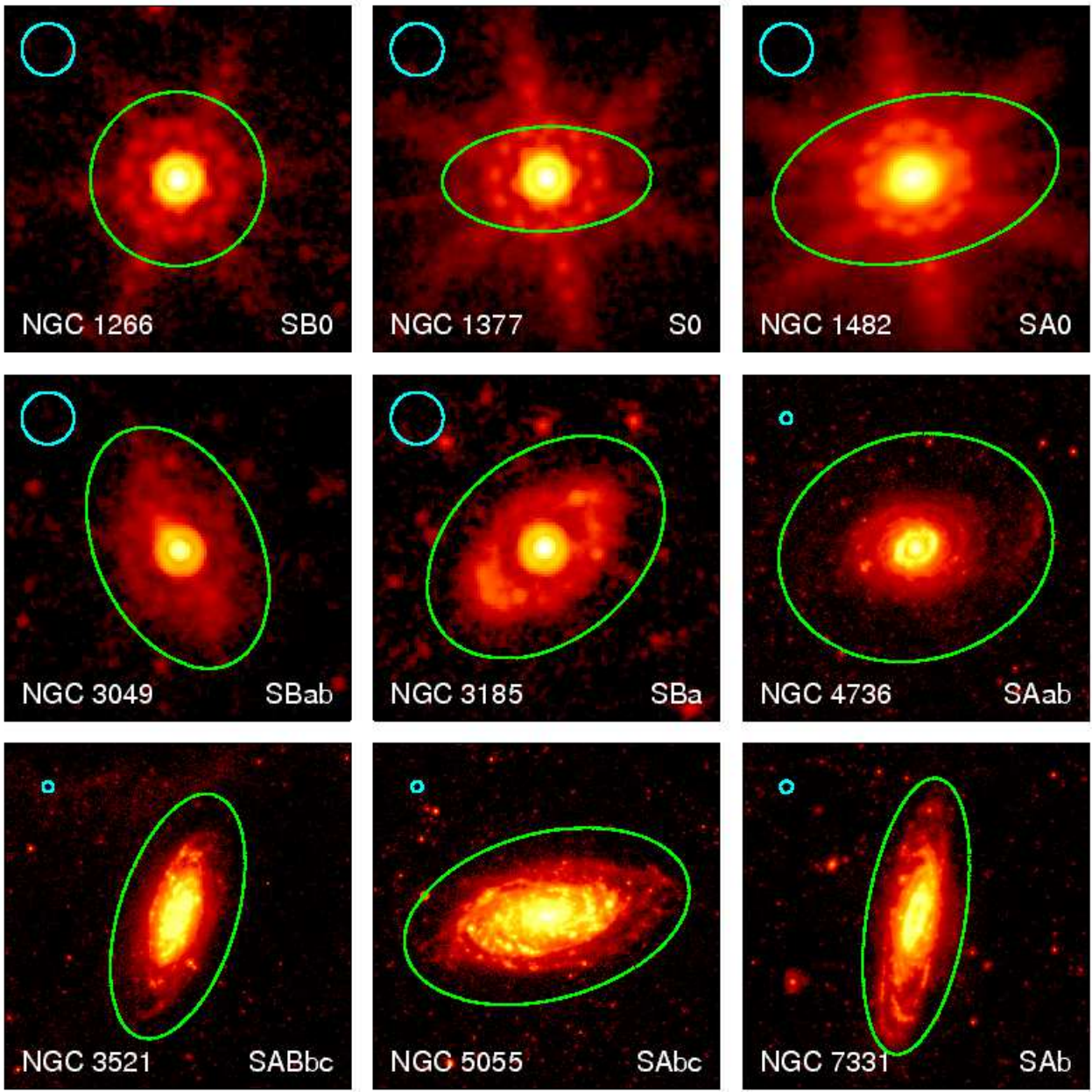}
\caption{24~$\mu$m images of galaxies with typical morphological
parameters for their Hubble types (see
Section~\ref{s_morphparam_rep}).  Each row contains galaxies with
similar Hubble types.  The images are ordered from E-S0 in the top row
to Sdm-Im in the bottom row.  The Hubble type of each galaxy (from
RC3) is listed in the lower corner of each image.  The D$_{25}$
isophote from RC3 is overplotted on the data.  An 18~arcsec
circle representing three times the FWHM of the 24~$\mu$m data is
plotted in the upper left corner of the images.  Note that most of the
extended structure in the NGC~1266, NGC~1377, and NGC~1482 images are
mostly the PSF of the unresolved or marginally resolved central
sources, and some of the extended structure in the NGC~3049 image is
also related to the PSF of the central source.}
\label{f_represent} 
\end{figure*}

\begin{figure*}
\epsfig{file=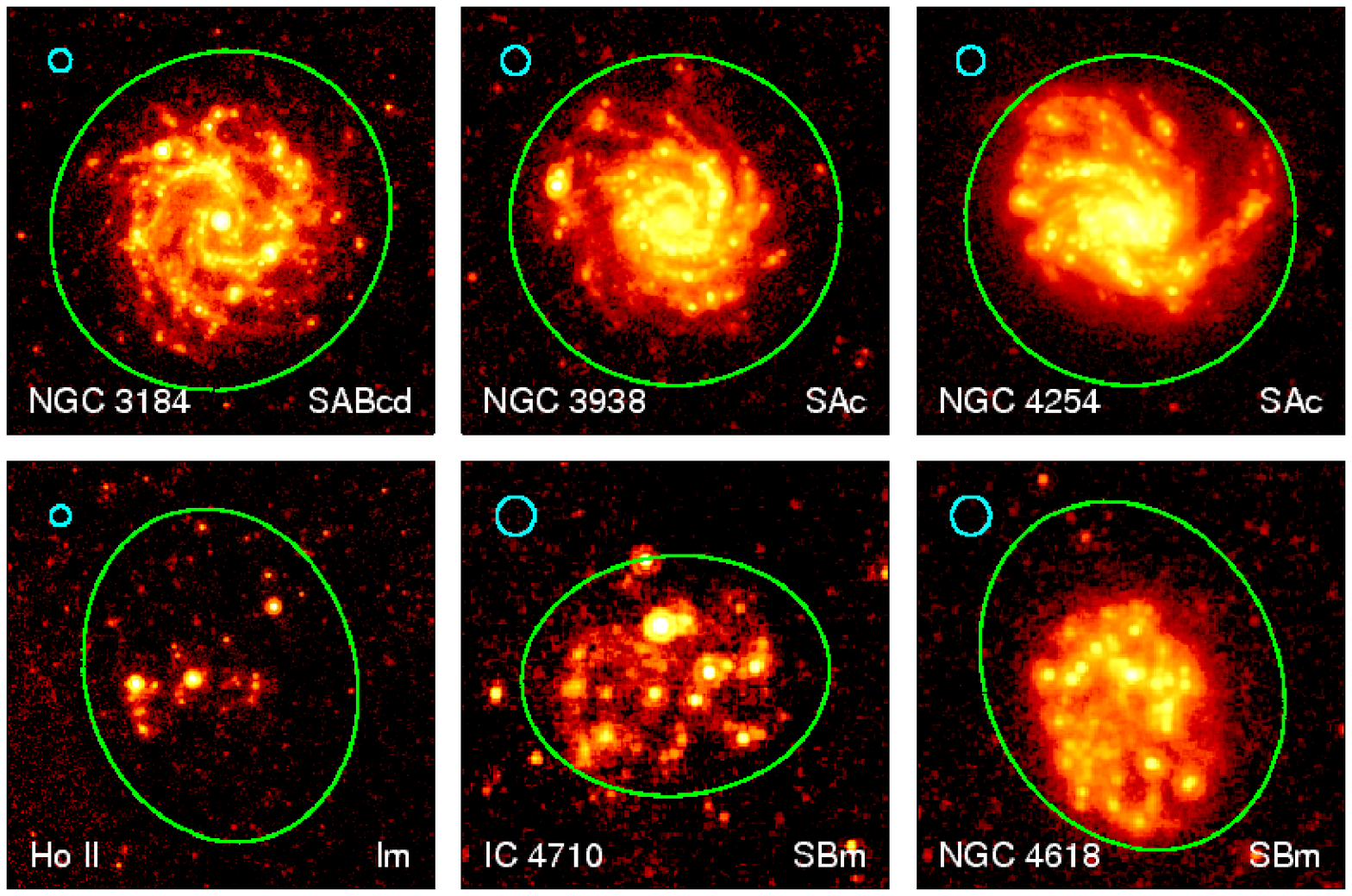}
\contcaption{}
\end{figure*}

The galaxies in Figure~\ref{f_represent} qualitatively illustrate the
variations in the distribution of 24~$\mu$m emission that were found
quantitatively in Section~\ref{s_morphparam_tanal}.  The 24~$\mu$m
emission from elliptical and S0 galaxies (when present) is generally
from a central, point-like or compact source.  In S0/a-Sab galaxies,
the nucleus is still the predominant source of emission, and the dust
outside the nucleus is generally axisymmetric.  Often, some of the
dust is located in ring-like structures as seen in NGC~3185 and
NGC~4736.  The 24~$\mu$m emission is relatively compact compared to
the optical disc of the galaxy.  Progressing through Sb-Sbc to Sc-Sd
galaxies, 24~$\mu$m emission from the disc becomes relatively strong
compared to the nucleus.  In many Sc-Sd galaxies, the nucleus is often
no longer the brightest source of emission, and the dust emission
appears only loosely organized.  Finally, in Sdm-Im galaxies, the
24~$\mu$m emission appears amorphous, although, as noted in the
previous section, the distribution of 24~$\mu$m emission in Sdm-Im
galaxies is quite varied.

\subsection{Galaxies with Peculiar 24~$\mu$m Morphologies}
    \label{s_morphparam_pec}

\begin{figure*}
\epsfig{file=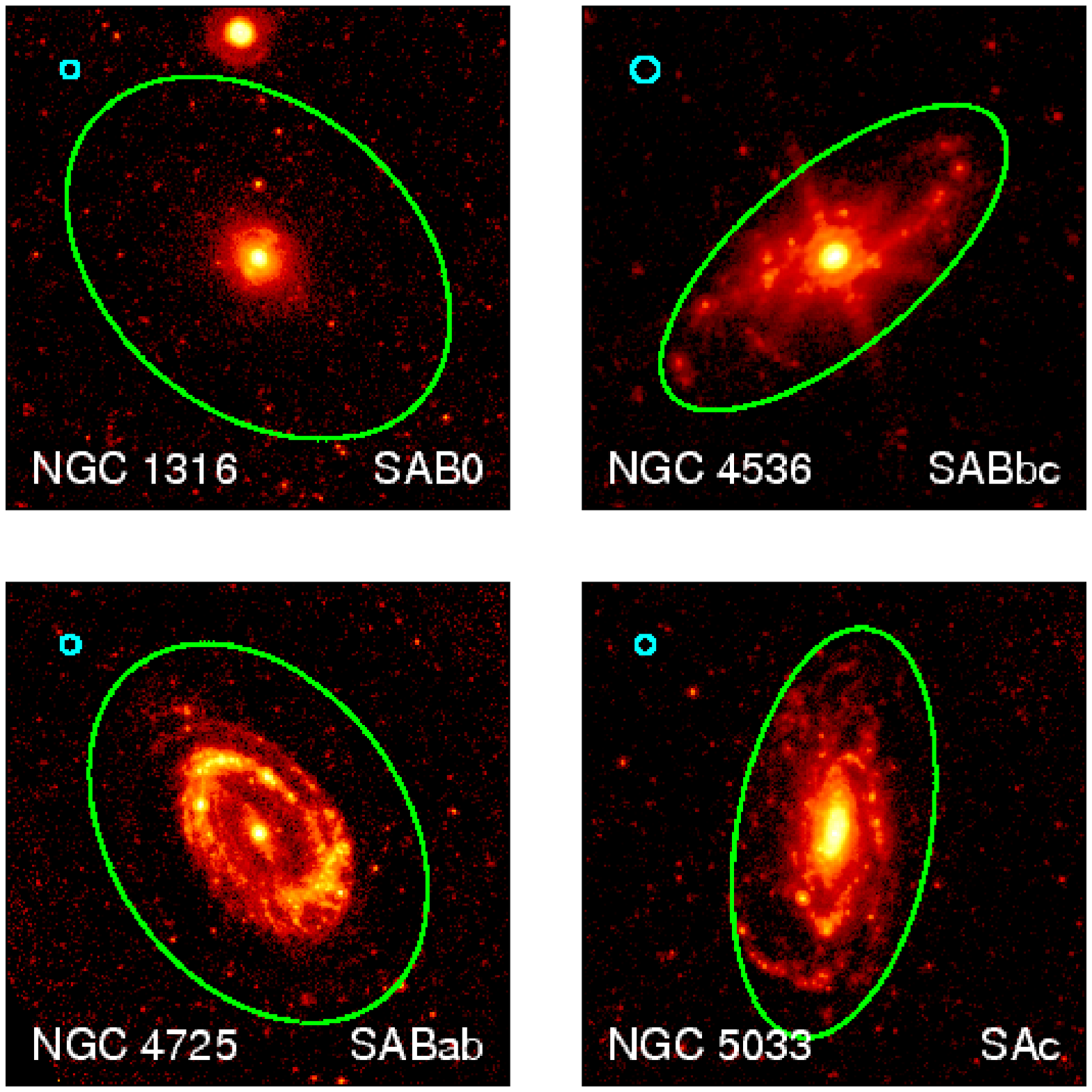}
\caption{24~$\mu$m images of galaxies with peculiar morphological
parameters for their Hubble types (see
Section~\ref{s_morphparam_pec}).  The images are ordered
alphanumerically.  See Figure~\ref{f_represent} for information on the
overplotted lines and the Hubble type.  In the image of NGC~4536, the
spike-like features perpendicular to the major axis of the galaxy and
the central ring structure is part of PSF of the marginally-resolved
central starburst.}
\label{f_peculiar}
\end{figure*}

The outliers from the trends in the 24~$\mu$m parameters with Hubble
type are also worth noting simply because they show how AGN activity,
starburst activity, or dynamical processes can affect either dust
heating or the distribution of dust within the galaxies.  Four
examples of notable peculiar sources (where the 24~$\mu$m
morphological parameters differ from the median values by a large
number of standard deviations) are listed below, with their 24~$\mu$m
images shown in alphanumeric order in Figure~\ref{f_peculiar}.

NGC~1316 (Fornax A).  Compared to other E-S0 galaxies, the 24~$\mu$m
emission is unusually centrally concentrated and unusually asymmetric.
This is exceptional because the 24~$\mu$m emission in E-S0 galaxies is
already centrally concentrated compared to other Hubble types.  This
galaxy is a radio galaxy \citep{w61} that has undergone multiple
interactions \citep{s80}.  The central far-infrared emission from this
galaxy is probably strongly enhanced by the AGN activity.  Moreover,
the interactions have triggered gas infall into the nucleus
\citep{s80}, which would also make the 24~$\mu$m appear centrally
concentrated.  These interactions may also explain the asymmetric
appearance of the galaxy at 24~$\mu$m.  Note that AGN may also be
responsible for producing centrally concentrated 24~$\mu$m emission in
some of the other E-S0 galaxies as well.

NGC~4536.  The 24~$\mu$m emission is unusually centrally concentrated
in NGC~4536 compared to other Sb-Sbc galaxies as demonstrated by the
low $\overline{M}_{20}$ and $\overline{R}_{eff}$ values.  This galaxy
is an example of a late-type spiral galaxy with strong nuclear
starburst activity \citep[e.g.][]{phm88,tdw93}, which is why the
infrared emission appears very centrally concentrated.  Other galaxies
that contain strong nuclear star formation activity, including
NGC~4631 \citep[e.g.][]{rkh92} and NGC~6946
\citep[e.g.][]{th83,dy93,tdw93}, also appear unusually centrally
concentrated at 24~$\mu$m.

NGC~4725.  A significant amount of the 24~$\mu$m emission from
NGC~4725 is associated with the galaxy's asymmetric outer ring
\citep[e.g.][]{b88}.  Consequently, the 24~$\mu$m emission from the
galaxy is the most extended and asymmetric of all the S0/a-Sab galaxy
in the sample.  NGC~1291 also contains a dusty outer ring structure
that also makes it appear unusually extended at 24~$\mu$m, although it
does not appear as extended or asymmetric as NGC~4725.

NGC~5033.  This galaxy has a Seyfert 1.5 nucleus \citep{hfs97b}.  As a
consequence, the 24~$\mu$m emission appears to be enhanced by the AGN
activity, so the 24~$\mu$m emission is more centrally concentrated
than the average Sc-Sd galaxy.  Note that some of the other Seyfert
galaxies in this sample, such as NGC~3185 \citep{hfs97b}, are also
very compact.

\section{Variations in 24~$\mu$m Parameters between Barred and Unbarred
    Spiral Galaxies} \label{s_morphparam_bar}

The effects of bars on the ISM of galaxies has been well documented.
Both observations \citep[e.g.][]{sois99,svrtt05} and theoretical
models \citep[e.g.][]{a92,fb93} have demonstrated that bars drive gas
into the centres of spiral galaxies.  Consequently, bars may enhance
nuclear star formation activity \citep[e.g.][]{hetal96, hfs97a,
retal01}, although some observational studies have demonstrated that
not all galaxies with bars have enhanced nuclear star formation
\citep[e.g.][]{mf97, bj04}.  Nonetheless, if bars do enhance nuclear
star formation, then the 24~$\mu$m morphological parameters should
demonstrate that the 24~$\mu$m emission in barred galaxies is more
compact and possibly more peaked compared to unbarred galaxies.

\begin{figure*}
\epsfig{file=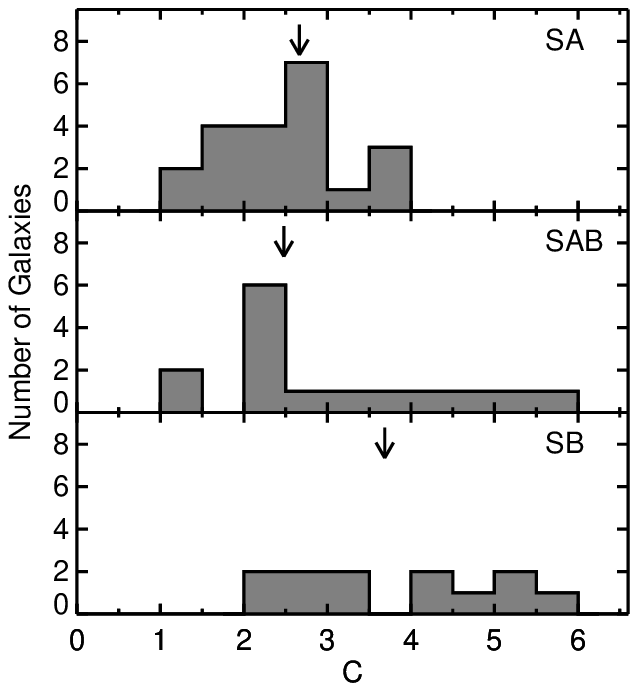, height=70mm}
\epsfig{file=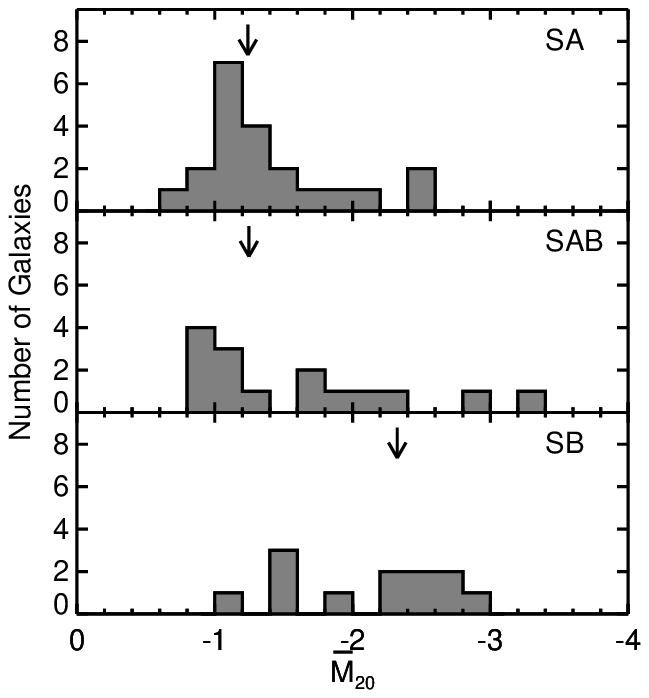, height=70mm}
\epsfig{file=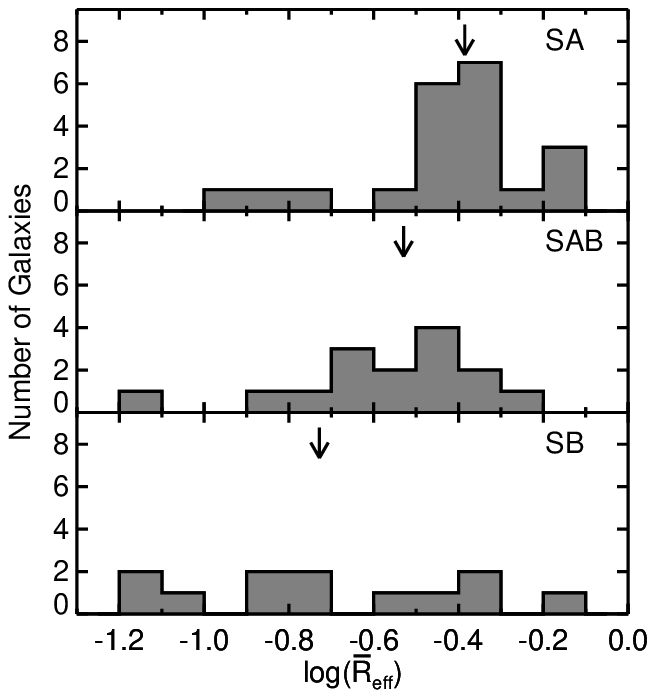, height=70mm}
\epsfig{file=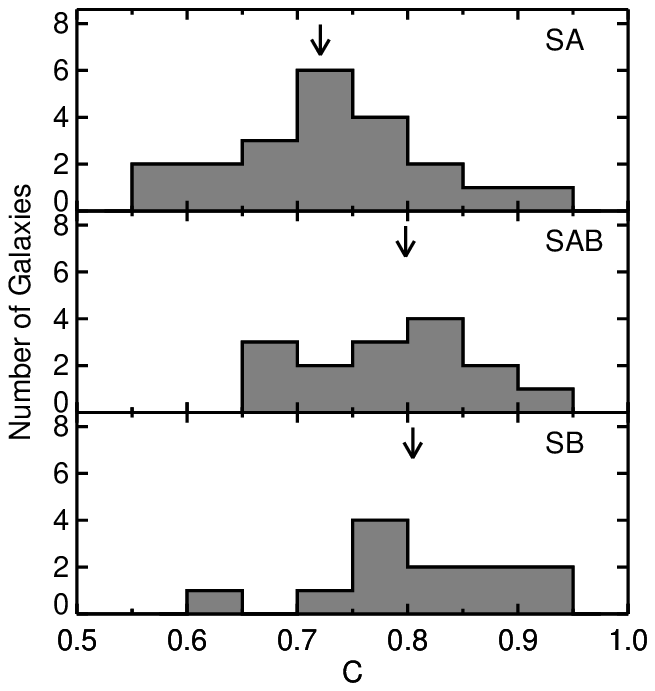, height=70mm}
\caption{Histograms of the 24~$\mu$m $C$, $\overline{M}_{20}$, and
$\log(\overline{R}_{eff})$ parameters for all S0/a-Sd galaxies sorted
according to bar type.  The $C$ and lower $\overline{M}_{20}$
parameters are plotted so that more concentrated (higher $C$, lower
$\overline{M}_{20}$) values appear on the right.  The
$\log(\overline{R}_{eff})$ parameter is plotted so that more extended
(larger $\log(\overline{R}_{eff})$ values fall on the right.  The
arrow indicates the median value for each subgroup (with the average
of the two central values used to calculate the median in subgroups
with an even number of members, which is why the median does not
appear to correspond to any galaxy in one of the panels).  See
Section~\ref{s_morphparam_bar} for discussion.}
\label{f_param_vs_bar}
\end{figure*}

To examine the effects of bars, we compared the 24~$\mu$m
morphological parameters of unbarred (SA), weakly barred (SAB), and
strongly barred (SB) galaxies.  Comparison are made for all spiral
(S0/a-Sd) galaxies and for subsets of spiral galaxies divided
according to their Hubble type.  Table~\ref{t_param24_bar} shows the
median and standard deviations of the $C$, $\overline{M}_{20}$,
$\overline{R}_{eff}$, and $G$ parameters of SA, SAB, and SB
galaxies, and Table~\ref{t_kstest_bar} shows the results from the K-S
test when applied to subsets of SA and SB galaxies.  Additionally,
Figure~\ref{f_param_vs_bar} shows histograms of the $C$,
$\overline{M}_{20}$, and $\overline{R}_{eff}$ parameters
for S0/a-Sd galaxies separated according to their bar type.

\begin{table*}
\centering
\begin{minipage}{127mm}
\renewcommand{\thefootnote}{\alph{footnote}}
\caption{Statistics on 24~$\mu$m Morphological Parameters for
    Different Bar Types}
\label{t_param24_bar}
\begin{tabular}{@{}lcccccc@{}}
\hline
    Hubble &
    Bar &
    Number of &
    $C$\footnotemark[1] &
    $\overline{M}_{20}$\footnotemark[1] &
    $\log(\overline{R}_{eff})$\footnotemark[1] &
    $G$\footnotemark[1] \\
    Type &
    Type &
    Galaxies &
    &
    &
    &
    \\
\hline
S0/a-Sd & SA & 21&
    $2.66 \pm 0.16$ &
    $-1.24 \pm 0.11$ &    
    $-0.39 \pm 0.05$ &
    $0.72 \pm 0.02$ \\
 & SAB & 15&
    $2.5 \pm 0.4$ &
    $-1.25 \pm 0.19$ &    
    $-0.53 \pm 0.06$ &
    $0.80 \pm 0.02$ \\
 & SB & 12&
    $4.1 \pm 0.4$ &
    $-2.31 \pm 0.17$ &    
    $-0.70 \pm 0.09$ &
    $0.81 \pm 0.02$ \\
S0/a-Sab & SA & 7 &
    $2.8 \pm 0.2$ &
    $-1.7 \pm 0.2$ &    
    $-0.39 \pm 0.10$ &
    $0.72 \pm 0.04$ \\
 & SAB & 3 &
    $2.4 \pm 1.2$ &
    $-1.7 \pm 0.7$ &    
    $-0.70 \pm 0.15$ &
    $0.80 \pm 0.06$ \\
 & SB & 6 &
    $4.5 \pm 0.5$ &
    $-2.47 \pm 0.19$ &    
    $-0.70 \pm 0.11$ &
    $0.85 \pm 0.05$ \\
Sb-Sbc & SA & 4 &
    $2.8 \pm 0.3$ &
    $-1.24 \pm 0.10$ &    
    $-0.35 \pm 0.03$ &
    $0.78 \pm 0.04$ \\
 & SAB & 7 &
    $3.1 \pm 0.5$ &
    $-1.7 \pm 0.3$ &    
    $-0.56 \pm 0.09$ &
    $0.83 \pm 0.02$ \\
 & SB & 2 &
    $5.2 \pm 0.6$ &
    $-2.34 \pm 0.16$ &    
    $-1.1282 \pm 0.019$ &
    $0.89 \pm 0.03$ \\
Sc-Sd & SA & 10 &
    $2.2 \pm 0.2$ &
    $-1.09 \pm 0.16$ &    
    $-0.33 \pm 0.06$ &
    $0.72 \pm 0.03$ \\
 & SAB & 5 &
    $2.5 \pm 0.7$ &
    $-1.1 \pm 0.2$ &    
    $-0.42 \pm 0.06$ &
    $0.77 \pm 0.03$ \\
 & SB & 4 &
    $2.8 \pm 0.7$ &
    $-1.41 \pm 0.14$ &    
    $-0.37 \pm 0.13$ &
    $0.797 \pm 0.017$ \\
\hline
\end{tabular}
$^a$ These are median values and standard deviations of the means.
\end{minipage}
\end{table*}

\begin{table}
\begin{center}
\renewcommand{\thefootnote}{\alph{footnote}}
\caption{Results on Applying K-S Test to 24~$\mu$m Morphological
    Parameters for SA and SB Galaxies}
\label{t_kstest_bar}
\begin{tabular}{@{}ccc@{}}
\hline
    Hubble Type &
    Parameter &    Probability\footnotemark[1] \\
\hline
S0/a-Sd &   $C$ &                       0.018\\
 &  $\overline{M}_{20}$ &               0.0060\\
 &  $\log(\overline{R}_{eff})$ &        0.0095\\
 &  $G$ &                               0.015\\
S0/a-Sab &  $C$ &                       0.19\\
 &  $\overline{M}_{20}$ &               0.048\\
 &  $\log(\overline{R}_{eff})$ &        0.19\\
 &  $G$ &                               0.54\\
Sb-Sbc &    $C$ &                       \footnotemark[2]\\
 &  $\overline{M}_{20}$ &               \footnotemark[2]\\
 &  $\log(\overline{R}_{eff})$ &        \footnotemark[2]\\
 &  $G$ &                               \footnotemark[2]\\
Sc-Sd &     $C$ &                       0.16\\
 &  $\overline{M}_{20}$ &               0.24\\
 &  $\log(\overline{R}_{eff})$ &        0.90\\
 &  $G$ &                               0.064\\
\hline
\end{tabular}
\end{center}
$^a$ This is the probability that the tested parameter for the SA and
     SB data have the same distribution.  A probability of $\sim0.01$
     is interpreted as having the same significance as a $3\sigma$
     difference between the median values for the S0/a-Sab and Sc-Sd
     galaxies. \\
$^b$ The K-S test results are not meaningful unless each dataset
     contains at least four values.  The K-S test would compare a set
     of 4 values to a set of 2 values in this particular case, so
     these probabilities would have no meaningful interpretation and
     are not reported here.
\end{table}
\renewcommand{\thefootnote}{\arabic{footnote}}

These results generally demonstrate a difference between SA and SB
galaxies, with SB galaxies having more centrally concentrated
24~$\mu$m emission as seen in higher $C$, lower $\overline{M}_{20}$,
and lower $\overline{R}_{eff}$ parameters and more emission located in
peaks as seen in the higher $G$ parameter.  However, the trends are
only significant at the $3\sigma$ level when all spiral galaxies (all
S0/a-Sd galaxies) are used in the analysis.  Since the SA galaxies for
this sample tend to be a slightly later type than the SB galaxies for
this sample, the results from comparing all unbarred spiral galaxies
to all barred spiral galaxies may be partly biased by trends along the
Hubble sequence (although $G$ should exhibit this bias).  When subsets
of spiral galaxies (only S0/a-Sab, only Sb-Sbc, or only Sc-Sd
galaxies) are examined, SA galaxies may still be differentiated from
SB galaxies, but the results are not as statistically significant.
Unfortunately, when the galaxies in this sample are divided according
to both Hubble type and bar type, the resulting subsets are relatively
small; all subsets contain fewer than 10 galaxies, and some subsets
contain fewer than 5 galaxies.  We conclude that bars may play a role
in changing the 24~$\mu$m morphologies of nearby galaxies, which would
be consistent with previous observations of barred spiral galaxies,
but that a larger sample of spiral galaxies is needed to examine these
trends in more detail.

Although the analysis is not entirely conclusive in detecting
differences between barred and unbarred galaxies, we note that the
results do show differences between barred and unbarred galaxies in
both the early-type spiral galaxy subset and the late-type spiral
galaxy subset.  Most recent observations show that bars only enhance
nuclear star formation activity in early-type spiral galaxies
\citep{hfs97a, retal01}, which implies that bars should only affect
the 24~$\mu$m morphologies of early-type spiral galaxies.  However, we
should emphasize that comparing the 24~$\mu$m morphological parameters
to measurements of nuclear star formation activity may not be
appropriate.  Although the 24~$\mu$m morphologies show that the
distribution of star formation itself changes, it cannot be used to
show changes in the relative strength of star formation activity
(i.e. the ratio of current to past star formation activity, or the
star formation activity per unit total stellar mass).  A comparison of
the morphological parameters to measurements of the relative star
formation activity is needed to better understand the trends shown by
the morphological parameters.

\section{Discussion} \label{s_discuss}

\subsection{Comparison to Results on Stellar Morphologies} 
    \label{s_discuss_compstellar}

We compared the 24~$\mu$m morphological parameters for the data in
Table~\ref{t_param24} to the data for 3.6~$\mu$m morphological
parameters in Table~\ref{t_paramirac1} and to the R-band morphological
parameters given in \citet{c03}.  In general, the 24~$\mu$m parameters
differ significantly from the 3.6~$\mu$m and R-band parameters.  For
an individual Hubble type, the 24~$\mu$m emission appears more
extended then the 3.6~$\mu$m or R-band emission, as seen by the lower
$C$, higher $\overline{M}_{20}$, and higher $\log(\overline{R}_{eff})$
parameters.  The 24~$\mu$m emission also appears more asymmetric, as
indicated by the higher $A$ parameters, and more of the 24~$\mu$m
emission appears to originate in one or a few bright peaks, as
exhibited by the higher $G$ parameter.  (Note, however, that the
3.6~$\mu$m $C$ and $A$ parameters in Table~\ref{t_paramirac1} are
comparable to the same R-band parameters given by \citet{c03}.)

\begin{figure*}
\epsfig{file=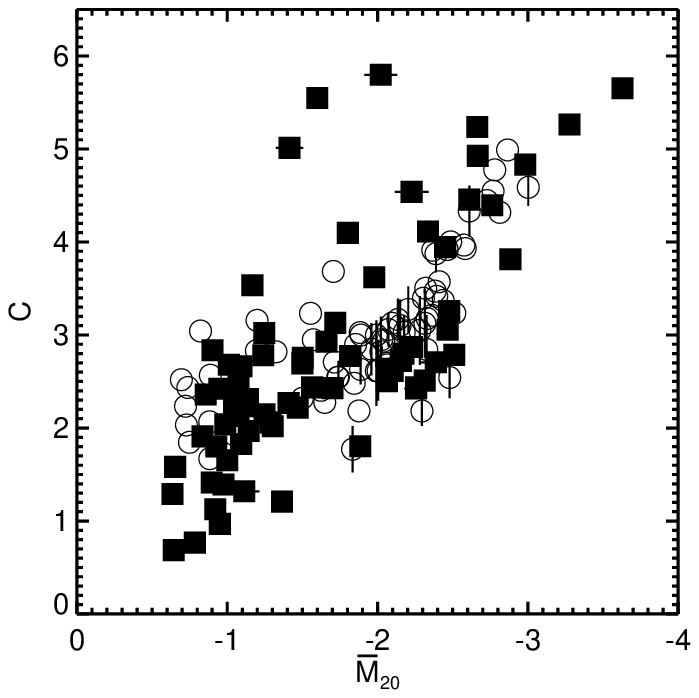, height=67mm}
\epsfig{file=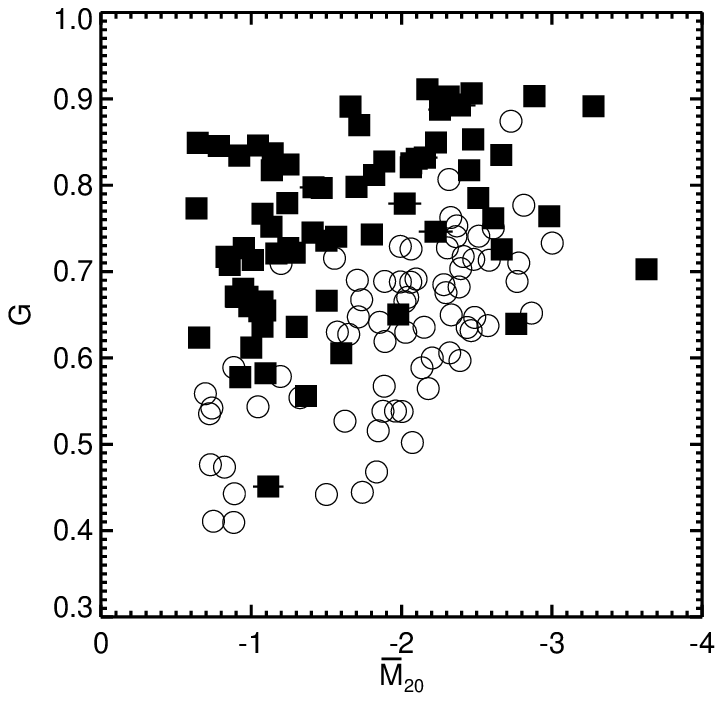, height=67mm}
\epsfig{file=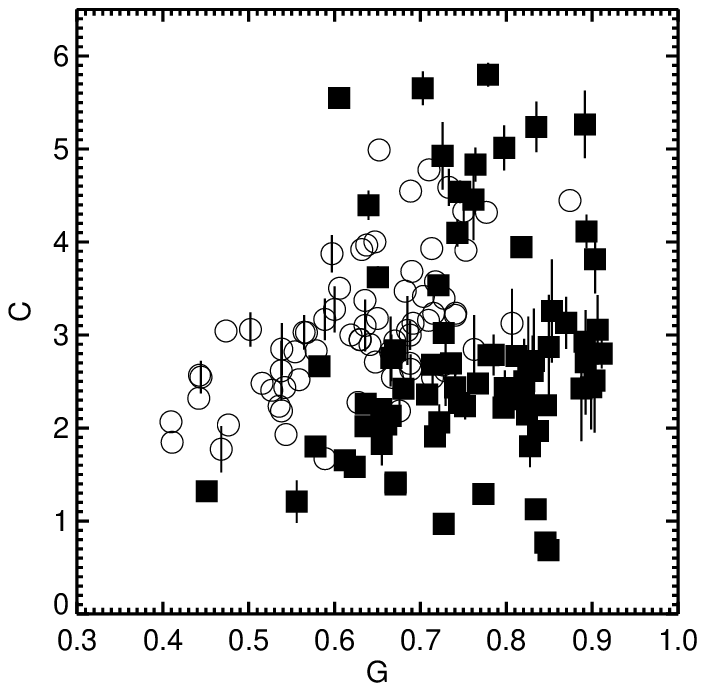, height=67mm}
\epsfig{file=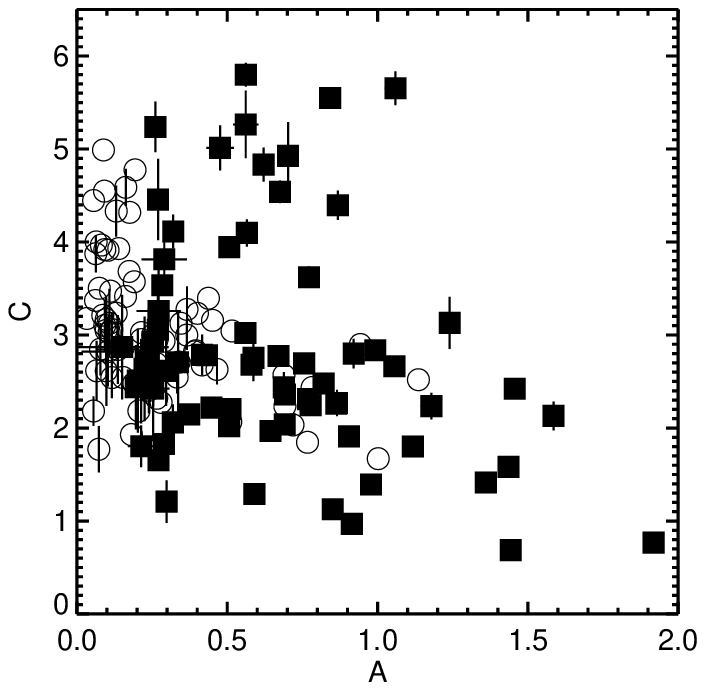, height=67mm}
\epsfig{file=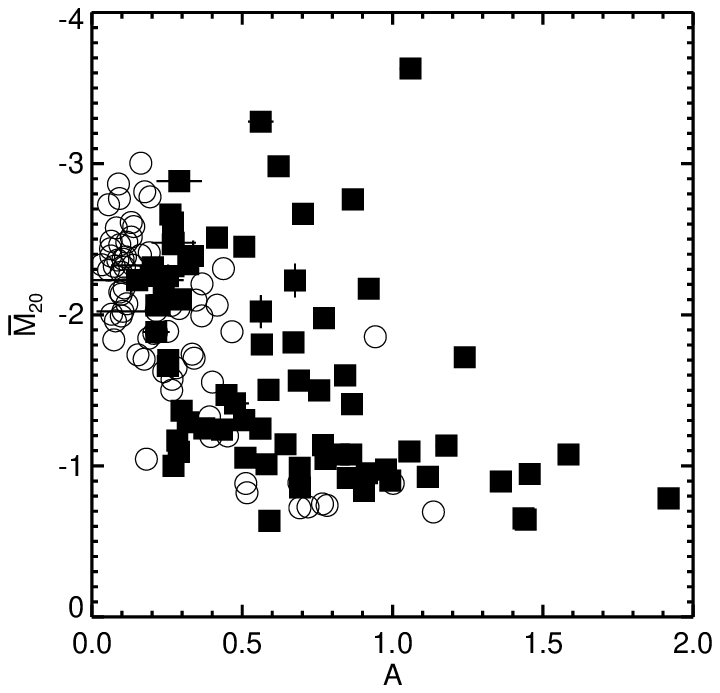, height=67mm}
\epsfig{file=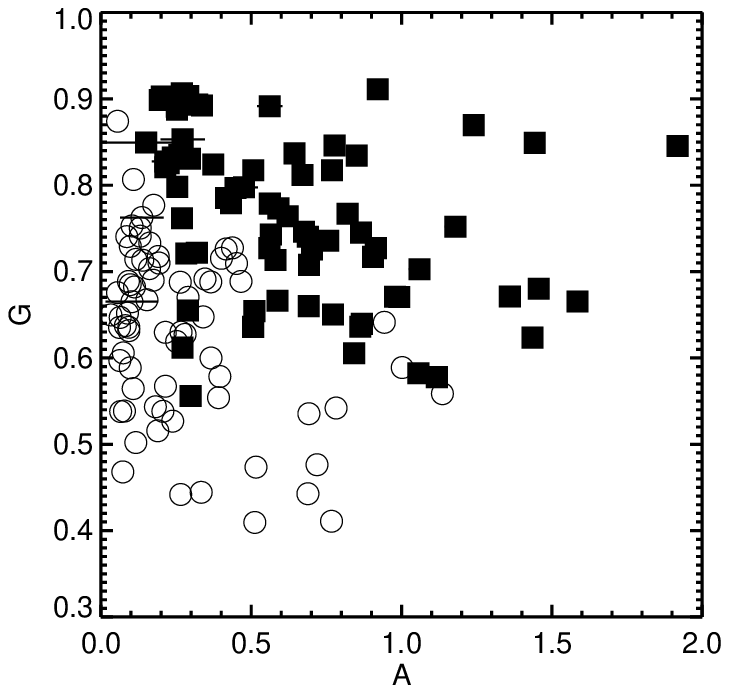, height=67mm}
\caption{Plots of the 3.6~$\mu$m (open circles) and 24~$\mu$m (filled
squares) morphological parameters versus each other.  These plots are
meant to be directly compared with Figures 10-14 in \citet{lpm04}
although the scales of the x- and y-axes are slightly different.  The
plot of $C$ versus $A$ may also be compared to Figure 15 of
\citet{c03}, although Conselice plots the data differently.  Note that
Ho~I, which has a 24~$\mu$m asymmetry of $3.28 \pm 0.11$, falls
outside some of the plots.  See Section~\ref{s_discuss_compstellar}
for discussion.}
\label{f_paramcomp}
\end{figure*}

In Figure~\ref{f_paramcomp}, we show comparisons of the morphological
parameters to each other that may be directly compared with Figure~15
in \citet{c03} and Figures~10-14 in \citet{lpm04}.  The parameter
space occupied by the 3.6~$\mu$m SINGS data is similar to the
parameter space occupied by the R-band data in Conselice and Lotz et
al.  This is expected, as both the 3.6~$\mu$m wave band and the R-band
trace the starlight from all stars within the host galaxies, including
evolved red stars.  However, the parameter space occupied by the
24~$\mu$m morphological parameters for the SINGS galaxies do not
correspond to the same parameter space occupied by the 3.6~$\mu$m
morphology data for the SINGS galaxies or the R-band morphology data
for normal galaxies in \citet{c03} and \citet{lpm04}.  The offset
between the 3.6 and 24~$\mu$m parameters is most easily seen in the
plots of $\overline{M}_{20}$ versus $G$, $C$ versus $G$, and $A$
versus $G$.  However, note that the 24~$\mu$m $C$ and $A$ parameters
for the SINGS galaxies are comparable to the corresponding R-band
parameters for starbursts and mergers in \citet{c03}, and the
24~$\mu$m data occupy a similar location in parameter space as the
R-band data for ULIRGs in \citet{lpm04}.

Aside from showing the differences between the distribution of
starlight and 24~$\mu$m emission in nearby galaxies, these results
demonstrate that individual wave bands need to be calibrated at rest
wavelengths to determine the parameter spaces occupied by both normal
and merging galaxies.  It may also be desirable to demonstrate that
morphological parameters measured in wave bands that are sensitive to
star formation can still be used to separate mergers and starbursts
from normal galaxies, especially given that the 24~$\mu$m
morphological parameters for the galaxies in this paper statistically
match the R-band morphological parameters for starbursts and ULIRGs.

Despite the differences in the parameter values between the 24~$\mu$m
wave band and stellar wave bands, the 24~$\mu$m morphological
parameters show variations along the Hubble sequence that are similar
to those seen in the 3.6~$\mu$m morphological parameters as well as
the R-band parameters from \citet{c03}.  The morphological parameters
measured in stellar wave bands and at 24~$\mu$m show that elliptical
and early-type spiral galaxies are generally compact and symmetric
whereas late-type spiral galaxies are more extended and asymmetric.
Moreover, the differences in the 24~$\mu$m asymmetry between early-
and late-type spiral galaxies observed here does resemble the trends
in lopsidedness observed in near-infrared observations of nearby
galaxies by \citet{bcjp05}.  The $G$ parameter is the only parameter
where the trends seen at 24~$\mu$m differ from those seen in stellar
wave bands.  Slightly more 3.6~$\mu$m emission appears to originate
from one or a few bright peaks in early-type spiral galaxies when
compared to late-type spiral galaxies, but the same fraction of
24~$\mu$m emission is located in bright peaks in both early- and
late-type spiral galaxies.  Nonetheless, we conclude that the
variations in stellar morphologies along the Hubble sequence are
interrelated with the dust morphologies and implicitly with the
morphologies of the ISM and the distribution of star formation
regions.

\subsection{Comparison to Results on the Distribution of the ISM and Star 
Formation}

The observed trends in the central concentration and spatial extent of
the 24~$\mu$m dust emission along the Hubble sequence are similar to
those observed at 12 and 60~$\mu$m by \citet{betal02}, at 850~$\mu$m
by \citet{tacgde04}, and at 8~$\mu$m by \citet{pafw04}.  However, the
results from the analysis in this paper illustrate the trend more
robustly than these previous studies; the analysis in this paper is
quantitative instead of qualitative, the images used in this paper
completely cover the optical discs of all the galaxies, and the sample
in this paper contains several galaxies of each Hubble type.
Interestingly, the same variations in the spatial distribution of dust
emission are found regardless of the wave band used. This implies that
a single wave band can be used to roughly characterize the spatial
extent of all the dust emission on global scales, although colour
variations within galaxies, such as those observed by \citet{detal05},
\citet{ckbetal05}, \citet{petal06}, and \citet{betal06b}, indicate
that a single wave band cannot be used to accurately trace the total
dust emission on kpc scales.

The weak relationship between morphology and the effective radius of
the 24~$\mu$m emission is similar to the trend observed with molecular
gas by \citet{yetal95}, which demonstrated that molecular gas was
found in more central concentrations in early-type galaxies and in
wider distributions in Sc galaxies.  This shows that, for Sa-Sc spiral
galaxies, variations in the distribution of 24~$\mu$m emission as a
function of morphology are related to the variations in the
distribution of molecular gas that fuels star formation.  The
similarities of the trends is somewhat surprising, as both 24~$\mu$m
and CO emission are expected to trace gas mass in the ISM, but the
24~$\mu$m emission is also expected to be highly sensitive to
variations in the illuminating radiation field \citep{dhcsk01}.
However, since the 24~$\mu$m wave band does trace star formation
\citep{ckbetal05, petal06} and since the Schmidt law states that star
formation is directly related to gas density \citep{k98b}, the
relation between the distribution of CO and 24~$\mu$m emission does
not seem too extraordinary.

However, \citet{yetal95} found that CO was more centrally concentrated
in Sd and later types of galaxies than in Sc galaxies, whereas we find
the opposite for dust emission.  The reason for the difference between
the Young et al.  results and our results for Sd and later galaxies
could be related to three phenomena that would make the CO emission
too weak to detect outside the centres of the galaxies.  First, the CO
line emission may be low in regions with low metallicities
\citep[e.g.]{tks98}.  Second, the ratio of molecular to atomic gas
varies as a function of Hubble type \citep[e.g.][]{yk89}, so less
molecular gas and hence less CO emission may be present.  Therefore,
the CO emission in the late-type galaxies of the Young et al. data may
only be detectable in the centres.  The dust emission, however, may be
related to the total gas content in these very late-type galaxies and
may be less sensitive to metallicity effects, so it will appear more
spatially extended than the CO emission.

We note that the exponential scale lengths of the H{\small I} gas
normalized by the optical scale lengths in the sample used by
\citet{tacgde04} did not vary with Hubble type, although their sample
for the H{\small I} analysis contained 2 S0-Sab galaxies.
Nonetheless, these results are contrary to the trends for 24~$\mu$m
dust emission found here and the trends for CO and dust emission found
elsewhere.  Additional studies of H{\small I} are needed to further
understand how its distribution varies as a function of Hubble type
and its interrelation with dust emission at different wavelengths.

Finally, we observe that the variations in the spatial extent of
24~$\mu$m emission with Hubble type are similar to the variations in
the radial distributions of H{\small II} regions observed by
\citet{hk83}.  However, the variations we observed in the spatial
extent of the 24~$\mu$m emission with Hubble type do not agree with
\citet{dghhc01}, who found no such variations in the ratio of the
spatial extent of H$\alpha$ and I-band emission among early- and
late-type cluster spiral galaxies.  The reason for this discrepancy is
unclear.  One possibility is that the differences in the spatial
distribution of star formation among cluster spiral galaxies are
relatively small compared to field spiral galaxies, so the Dale et
al. sample, comprised entirely of cluster galaxies, would show no
variations in the spatial extent of star formation, whereas the sample
here, which contains mostly field galaxies, would show such
variations.  The numbers of Virgo Cluster galaxies in our sample is
too small to accurately assess whether the spatial extent of 24~$\mu$m
emission in early- and late-type spiral galaxies is also invariant
within clusters.  The observed variations in the spatial extent of the
24~$\mu$m emission also appear to contradict the results from
\citet{khc06}, who found no such variations in the H$\alpha$/R-band
scale lengths among spiral galaxies.  Because of problems with stellar
continuum subtraction, continuum from AGN, and dust extinction,
Koopmann et al. needed to exclude the central regions of the galaxies
from their analysis.  The H$\alpha$ radial profiles measured by
Koopmann et al. therefore would not have been able to show variations
in the H$\alpha$ surface brightnesses in the centres of galaxies among
galaxies of different morphologies.  Although the 24~$\mu$m data are
affected by AGN, they are not as strongly affected by stellar
continuum or dust extinction problems, so they may show more accurate
variations in the spatial extent of star formation among galaxies of
different morphologies.

\subsection{Implications for Interpreting the Hubble Sequence}
    \label{s_discuss_hubbleseq}

\begin{figure*}
\epsfig{file=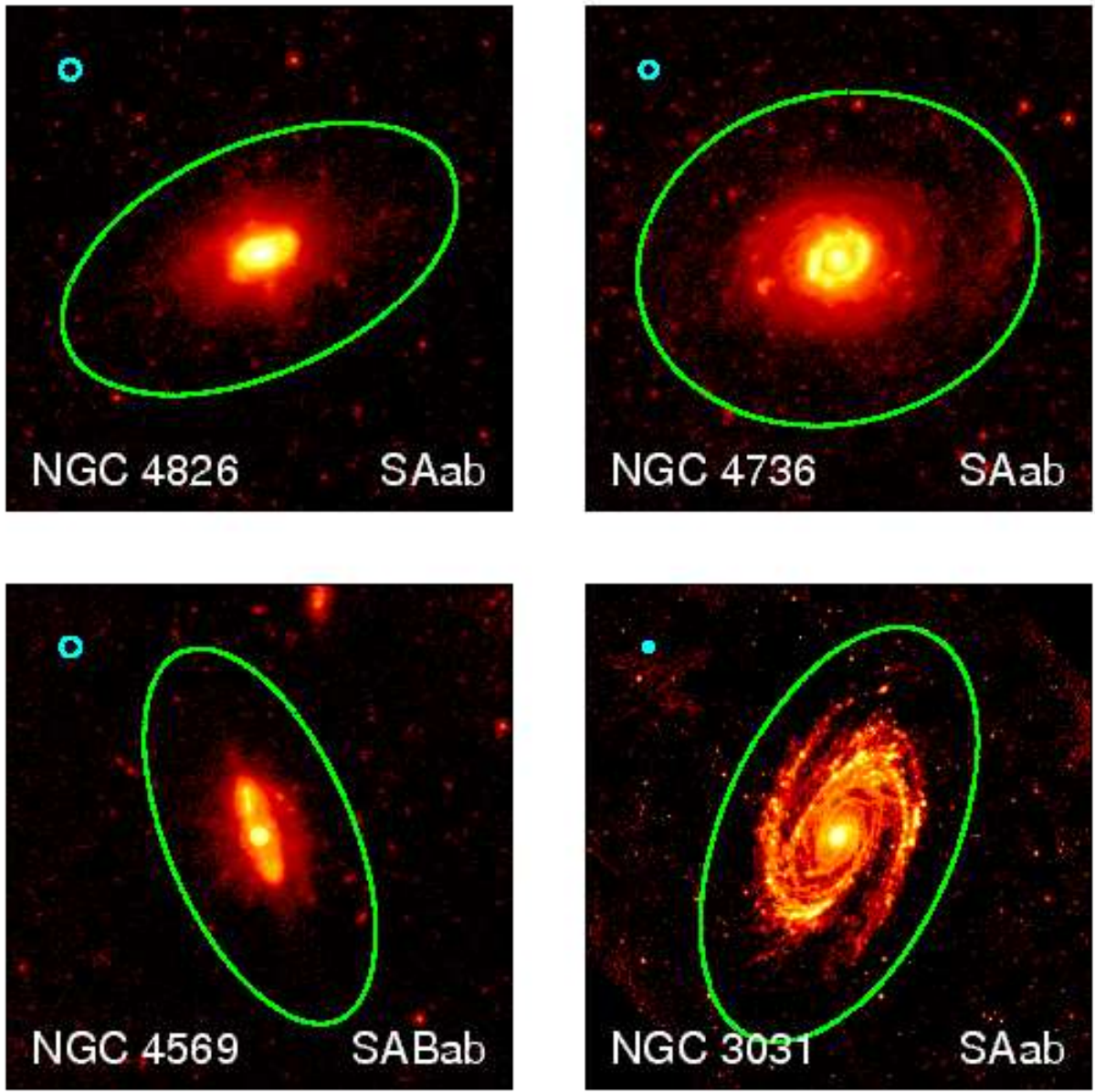}
\caption{24~$\mu$m images of four Sab galaxies.  See
Section~\ref{s_discuss_hubbleseq} for details.  See
Figure~\ref{f_represent} for information on the overplotted lines and
the Hubble type.}
\label{f_sabvar}
\end{figure*}

Trends in integrated star formation activity along the Hubble sequence
have been very well studied.  Most observational studies have shown
that global star formation varies significantly along the Hubble
sequence, with star formation per unit stellar mass increasing from
E-S0 through to Sd galaxies \citep[see][ for a review]{k98a}.
However, these trends appeared to be much weaker when examining the
nuclei of spiral galaxies.  The results on the 24~$\mu$m morphologies
here provides additional insights into why the variations are seen
primarily in the galaxies' discs and how the galaxies' morphologies
are tied to star formation in the discs.

A number of mechanisms have been proposed for building galaxies with
larger bulges or galaxies that appear to have larger bulges.  First,
inequal mass merger events may transform smaller galaxies with modest
bulge/disc ratios into larger galaxies with much larger bulge/disc
ratios \citep[e.g.][Sections 5.5 and 5.6]{s98}.  Second, pseudobulges
can be formed when bars or other dynamical mechanisms funnel gas into
the centres of galaxies \citep{kj-kr04}.  While not true bulges, these
pseudobulges may look like bulges when viewed face-on and may result
in galaxies being classified as early-type spiral galaxies.  Third, in
cluster environments, ram pressure stripping may remove gas from the
discs of spiral galaxies \citep{gg72}.  As the stellar populations
evolve, the discs, where star formation is repressed, become fainter
compared to the nuclei, where gas is still present and where star
formation continues.  Such galaxies would appear to have large
bulge/disc ratios and would be classified as early-type spiral
galaxies or S0 galaxies.

In each of the above scenarios, the changes in the distribution of the
ISM in the galaxies should be very similar.  In the merger scenario,
the gas will fall into the centre of the galaxy.  In the pseudobulge
scenario, the dynamical effects that create the pseudobulge also move
the ISM into the centres of the galaxies.  In the ram pressure
stripping scenario, the only remaining ISM in the galaxies is located
in the centres of the galaxies.  Overall, all of these mechanisms will
make the ISM appear more centralized in early-type spiral and S0
galaxies when compared to late-type spiral galaxies.

This is exactly what is seen in the 24~$\mu$m data presented here.
Disc galaxies with large bulges (S0-Sab galaxies) generally have more
centrally concentrated dust emission than disc galaxies with smaller
bulges (Sc-Sd galaxies).  Figure~\ref{f_represent} and the analysis in
Section~\ref{s_morphparam_ttrend} clearly illustrate this trend.  As
an additional demonstration of the effects of these evolutionary
processes, Figure~\ref{f_sabvar} shows four examples of Sab galaxies
that have undergone different evolutionary processes.  NGC~4826, a
galaxy that contains counterrotating gas in its disc \citep{bwk92}, is
a good example of an inequal-mass merger remnant.  Note the compact,
asymmetric structure in the 24~$\mu$m image.  NGC~4736 is a good
example of a galaxy with a pseudobulge \citep{kj-kr04}.
Although faint spiral arms are present in the outer disc, the nucleus
and inner ring are the two most prominent sources of 24~$\mu$m
emission within this galaxy, making it appear very compact.  NGC~4569,
which lies within the Virgo Cluster, is an example of a galaxy that
has undergone ram pressure stripping \citep{kr-kj04}.
Almost no emission originates from the disc of this galaxy.  Although
these three galaxies have evolved differently, the end result has been
the same: a relatively large optical bulge/disc ratio, and relatively
compact 24~$\mu$m emission.

In contrast to all of these, NGC~3031 is shown as an example of a Sab
field galaxy with a ``classical'' bulge \citep{kj-kr04}
that has not evidently undergone a recent merger (although it is
interacting with other nearby galaxies).  This galaxy is relatively
extended compared to most other S0/a-Sab galaxies, and it
qualitatively and quantitatively appears similar to many late-type
spiral galaxies at 24~$\mu$m.  Along with NGC~4725, it may be
representative of a subset of early-type spiral galaxies that do not
follow the same trends in 24~$\mu$m morphology as most other
early-type spiral galaxies.  Hence, the processes that are responsible
for creating the real or apparent large bulge-to-disc ratios in many
of the early-type spiral galaxies in this sample may not be
responsible for the formation of the large bulges in early-type spiral
galaxies like NGC~3031 and NGC~4725.

While acknowledging some exceptions, we may conclude that the
distribution of 24~$\mu$m emission within spiral galaxies is a
fundamental property linked to galaxies' Hubble classification.  The
results from the 24~$\mu$m data might also be indicative of the
variations in the distribution of the cool dust or the cool ISM
(neutral atomic and molecular hydrogen gas).  However, the 24~$\mu$m
emission is affected by dust heating as well as dust mass, so it is
imperfect as a tracer of gas or dust mass; additional research is
needed to determine the relation between morphological type and
variations in the distribution of other components of the ISM.

Nonetheless, if the spatial distribution of the 24~$\mu$m emission is
similar to that for cool gas and dust, then the variations observed
here demonstrate that the secular processes that increase the real or
apparent bulge/disc ratios of disc galaxies also change the
distribution of the ISM within the galaxies.  The ISM in the resulting
S0 and early-type spiral galaxies is very centrally concentrated.
Denied the fuel needed for star formation, the stellar populations in
the discs of early-type spiral galaxies will appear older than in
late-type spiral galaxies, contributing to the differences in relative
star formation activity summarized by \citet{k98a}.  The nuclei of
these early-type spiral galaxies, however, will still have the fuel
needed for star formation activity, which is why the nuclei of early-
and late-type galaxies are difficult to distinguish in terms of star
formation activity as discussed by \citet{k98a}.

\subsection{The Variation in the Distribution of Infrared
    Emission in Dwarf Galaxies} \label{s_discuss_irr}

Wide variations in the distribution of either star formation or the
ISM in dwarf galaxies have been observed previously.  For example,
\citet{pb03}, using optical data, found that star formation complexes
were non-uniformly distributed throughout individual dwarf irregular
galaxies.  \citet{he06} found that some galaxies have centres that are
bluer and more peaked than the centres of other dwarfs, demonstrating
that star formation in dwarfs is sometimes centralized and sometimes
diffuse.  \citet{bckks06} have found that H~{\small I} was also
distributed in an irregular and clumpy distribution.

The large range in the 24~$\mu$m morphologies of dwarf galaxies found
in this paper generally reflects these previous results from the
literature.  The general amorphous morphology of the dwarf galaxies
reflects the stochastic nature of star formation within these
galaxies, as has been noted for some of the individual objects in this
sample \citep[e.g.][]{cwbetal05}.  Under these circumstances, dust
emission may no longer be expected to trace star formation well, as
extinction may be variable across the galaxies \citep[e.g.]{cwbetal05}
and large reservoirs of gas and dust not associated with star
formation may be present in some galaxies \citep[e.g.][]{bckks06, wetal07}.

\section{Conclusions} \label{s_conclusions}

The main results are as follows:

1. Statistically significant variations are seen in the 24~$\mu$m
morphologies of nearby galaxies, and significant differences are seen
between S0/a-Sab and Sc-Sd galaxies.  Early-type galaxies generally
have symmetrical, centrally concentrated 24~$\mu$m emission.  Late-type
galaxies generally have asymmetric, extended 24~$\mu$m
emission.  The variations along the Hubble sequence can be attributed
to a number of mechanisms that transform late-type galaxies into
early-type galaxies, including mergers, pseudobulge formation, and
ram-pressure stripping in clusters.

2. The 24~$\mu$m morphologies of Sdm-Im galaxies are quite varied.
Some appear to be compact, point-like sources.  Others appear to be
highly extended and disorganized.  Additionally, we noted that some
dwarf galaxies were non-detections in one or more MIPS bands, implying
weak or negligible star formation activity.  This basically mirrors
other results on the distribution of star formation or the ISM in
nearby galaxies.  The variability in the 24~$\mu$m morphologies
reflects the stochastic nature of star formation within these
galaxies.

3. Bars may make the 24~$\mu$m morphologies appear more centrally
concentrated, thus confirming that bars play a role in enhancing
nuclear star formation activity.  However, the statistical
significance of the results based on this sample are weak.

4. The trends in the 24~$\mu$m morphologies along the Hubble sequence
mirror trends in morphological parameters measured in wave bands
dominated by stellar emission.  However, the 24~$\mu$m morphological
parameters appear similar to the R-band morphological parameters for
mergers and starbursts.  These results indicate that morphological
parameters measured in a wave band sensitive to star formation may not
correspond to parameters measured in a wave band that traces older
stellar populations.

5. The trends in the 24~$\mu$m morphologies along the Hubble sequence,
including the differences between S0/a-Sab and Sc-Sd galaxies, appear
similar to previously-observed trends in dust and molecular gas
emission.  However, the trends appear contrary to H{\small I}
observations of nearby galaxies, which show no variations in the
distribution of H{\small I} between early- and late-type spiral
galaxies.  Further observations are needed to understand the
variations in the distribution of the multiple constituents of the ISM
and their interrelation.

The implications of these results are extensive.  Any proposed galaxy
formation mechanisms that distinguish between early- and late-type
galaxies or any proposed mechanisms that transform late-type galaxies
into early-type galaxies must account for the variations in the
distribution of the star formation.  Moreover, models of the star
formation histories of nearby galaxies must account for the
differences in the location of star formation activity between early-
and late-type spiral galaxies.  In other words, valid models of spiral
galaxies with large bulges cannot be created simply by adding large
bulges with old stellar populations to star-forming discs but must
account for how the location of star formation varies with the
bulge-to-disc ratio.

We also note that the results here demonstrate that research based on
infrared observations of one galaxy should be used cautiously.  Such
studies are able to examine multiwavelength dust emission in more
detail and can provide detailed analyses on the nature of the
emission.  However, the applicability of the results on dust emission
for single galaxies to galaxies of different Hubble types may be
limited.  At least in terms of the distribution of 24~$\mu$m emission,
some galaxies are not even representative of galaxies with similar
Hubble types.  The distribution of 24~$\mu$m emission in NGC~3031, for
example, cannot be treated as representative of all Sab galaxies, and
NGC~6946 is not representative of all Scd galaxies.  If other galaxy
properties are linked to the spatial extent of the dust emission or at
least the 24~$\mu$m emission \citep[as suggested by][, for
example]{detal07}, then studies of individual objects like NGC~3031
and NGC~6946 might not be applicable to galaxies with similar Hubble
types.

While the sample used in this work was adequate for uncovering
morphological differences among nearby galaxies, a much larger sample
would provide more statistically robust results and may also provide
additional insights as to the circumstances where exceptions from
morphological trends occur.  An unbiased sample, such as a
distance-limited sample, may also produce more robust results than the
sample used here, as explained in Section~\ref{s_data_sample}.
Additionally, larger samples of both field and cluster galaxies may
demonstrate the extent to which interactions within clusters,
particularly ram-pressure stripping, alter the distribution of dust
emission or star formation within galaxies.

\section*{Acknowledgements}

We would like to acknowledge D. Hollenbach and the reviewer for their
helpful comments.  Support for this work, part of the Spitzer Space
Telescope Legacy Science Program, was provided by NASA through
contract 1224769 issued by the Jet Propulsion Laboratory, California
Institute of Technology under NASA contract 1407.

\appendix
\section{Finding the centre positions for measuring the
$\overline{M}_{20}$ and A parameters} \label{s_findcentre}

To find the central position for calculating $\overline{M}_{20}$ and
A, we adapted some of the procedures used by \citet{lpm04}.  To
measure the $\overline{M}_{20}$, they recommend shifting the centre
position for measuring radius until $M_{tot}$ is minimised.  To
measure A, they recommend shifting the centre of rotation for the
rotated image.

We measured $\overline{M}_{20}$ and A after shifting the centre
positions to a series of locations within a $13\times13$ grid, with
the centre of the grid corresponding to the central position given in
RC3 and the grid positions spaced by 0.5 pixels (0.75~arcsec) to a
maximum distance of 3 pixels (4.5~arcsec) from the RC3 position.  We
determined that when $M_{tot}$ or A were minimised for centre
positions further than 1~pixel (1.5~arcsec) from the RC3 position and
when the corrections to $\overline{M}_{20}$ and A were $\gtrsim5$\%,
the objects were inherently asymmetric.  In these situations, the
positions corresponding to these minima probably are not the centres
of the objects but are instead positions where asymmetric structures
may have less of an effect on the measured $\overline{M}_{20}$ and A.
Objects where $M_{tot}$ was minimised for centre locations within 1
pixel of the RC3 position and where the change in $\overline{M}_{20}$
or A were $\gtrsim5$ were compact, off-center sources.  The positions
corresponding to these minima are certainly the centres of these
objects.

We therefore determined that shifting the centre position away from
the position given by RC3 was only appropriate if a minimum was found
within 1 pixel.  Thus, when $M_{tot}$ or A were minimised when the
centre position was shifted up to 1 pixel of the RC3 position, we used
that position for measuring $\overline{M}_{20}$ or A.  However, when
$M_{tot}$ or A were minimised when the centre position was shifted
further than 1 pixel from the RC3 position, we used the RC3 position
itself as the centre position.

\label{lastpage}

\end{document}